\documentclass[12pt,a4paper]{article}

\usepackage{times}
\usepackage{jheppub}
\makeatletter
\def\@fpheader{\relax}
\makeatother
\numberwithin{figure}{section}
\usepackage{graphicx}
\def\inc#1{\includegraphics[scale=.35]{#1}}

\usepackage{framed}
\definecolor{shadecolor}{rgb}{0.90,0.90,0.90}

\def\Nequals#1{$\mathcal{N}{=}#1$}
\def\cQ{\mathcal{Q}}
\def\SO{\mathrm{SO}}
\def\SU{\mathrm{SU}}
\def\SL{\mathrm{SL}}

\def\U{\mathrm{U}}
\def\bC{\mathbb{C}}
\def\bH{\mathbb{H}}
\def\bZ{\mathbb{Z}}
\def\tr{\mathop{\mathrm{tr}}\nolimits}
\def\diag{\mathop{\mathrm{diag}}\nolimits}
\def\Sym{\mathop{\mathrm{Sym}}\nolimits}
\def\SYM{\mathop{\mathrm{SYM}}\nolimits}
\def\Coulomb{\mathop{\mathrm{Coulomb}}\nolimits}
\def\Higgs{\mathop{\mathrm{Higgs}}\nolimits}
\def\hkq{/\!/\!/}
\def\IR{\text{IR}}
\def\vev#1{\langle#1\rangle}

\newtheorem{fact_}{Fact}[section]
\newenvironment{fact}%
{\begin{fact_}\begin{shaded}}%
{\end{shaded}\end{fact_}}

\newtheorem{fact?_}[fact_]{Fact?}
\newenvironment{fact?}%
{\begin{fact?_}\begin{shaded}}%
{\end{shaded}\end{fact?_}}

\def\removevspaceafterequation{\setlength{\belowdisplayskip}{0pt} \setlength{\belowdisplayshortskip}{0pt}}
\let\oldtext\text
\def\text#1{\oldtext{\upshape #1}}

\begin{document}

\preprint{IPMU-15-0041, UT-15-10}
\title{A review of the $T_N$ theory and its cousins}
\author{Yuji Tachikawa}
\affiliation{
 Department of Physics, Faculty of Science, \\
 University of Tokyo,  Bunkyo-ku, Tokyo 133-0022, Japan and \\
 Kavli Institute for Physics and Mathematics of the Universe,  \\
 University of Tokyo,  Kashiwa, Chiba 277-8583, Japan}

\abstract{
The $T_N$ theory is a four-dimensional \Nequals2 superconformal field theory that has played a central role in the analysis of supersymmetric dualities in the last few years. The aim of this review is to collect known properties of the $T_N$ theory and its cousins  in one place as a quick reference. 
}

%\subjectindex{11.25.-w, 11.25.Tq, 11.25.Yb}
\maketitle
\newpage
\section{Introduction}
A textbook of four-dimensional (4d) quantum field theory usually starts the discussion from a classical Lagrangian which is later  quantized. 
In the recent years,  it is increasingly common to study 4d quantum field theories for which no useful classical Lagrangian is known.
Due to the lack of better terminology, let us call them \emph{non-Lagrangian} theories. Otherwise these theories are completely normal: it has a Hilbert space of states, a Hamiltonian generating time translation, operators supported at points, etc.

Twenty years ago only a few such theories were known, originally found by Minahan and Nemeschansky \cite{Minahan:1996fg,Minahan:1996cj}, and it might have been alright in those days to put them aside as mere curiosities that do not fit in the grand scheme of things. 
In a paper by Argyres and Seiberg \cite{Argyres:2007cn}, however, we learned that these theories of Minahan and Nemeschansky appear as essential ingredients to describe the S-dual descriptions of a few particular ordinary \Nequals2 supersymmetric gauge theories that have  classical Lagrangian descriptions.
Gaiotto then demonstrated in \cite{Gaiotto:2009we} that, in fact, such non-Lagrangian theories  almost always appear when we study the S-dual of  ordinary \Nequals2 supersymmetric gauge theories, and that the totality of the S-dual operations can only be elucidated in terms of these non-Lagrangian theories. 

The most important among the non-Lagrangian theories found in these works is the $T_N$ theory \cite{Gaiotto:2009gz}, a 4d \Nequals2 superconformal theory with $\SU(N)^3$ symmetry. 
In the purely 4d language, this theory can be introduced as a component in the S-dual of a certain quiver gauge theory, 
but more intrinsically, it is defined as the 4d limit of a compactification of the 6d \Nequals{(2,0)} theory of type $\SU(N)$ on a sphere with three full punctures. 
In general, the infrared limit of the compactification of the 6d \Nequals{(2,0)} theory on a Riemann surface with punctures is called a class S theory\footnote{The \emph{S} of the class S theory stands for \emph{six}, and the \emph{T} in the $T_N$ theory stands for the word \emph{theory}. The author thanks Davide Gaiotto for this important comment on the history of science.}, and the $T_N$ theory is the fundamental ingredient to understand class S theories.
Indeed, most of the newly found non-Lagrangian theories and most of the S-dualities among them are known to come  from various properties of the $T_N$ theory.

Due to this central role played by the $T_N$ theory, people devised various ways to obtain the properties of this theory, without relying on the classical Lagrangian description of this theory itself. 
For example, conformal and flavor central charges were studied in \cite{Gaiotto:2009gz,Bonelli:2009zp,Alday:2009qq,Chacaltana:2012zy}, 
various chiral ring relations were found in \cite{Benini:2009mz,Gadde:2013fma,Maruyoshi:2013hja,Hayashi:2014hfa},
and the superconformal indices have been intensively studied e.g.~in \cite{Gadde:2009kb,Gadde:2011ik,Gadde:2011uv,Gaiotto:2012xa}.
The properties of the $T_N$ theory is by now quite well understood, to the point that we can study a supersymmetry-breaking model that has the $T_N$ theory as an essential ingredient \cite{Maruyoshi:2013ega}.
There are various cousins of the $T_N$ theory, either by starting from a 6d \Nequals{(2,0)} theory of type $D$ or $E$, or by \emph{partially closing} the punctures of the $T_N$ theory. 
We now have an extensive series of papers \cite{Chacaltana:2010ks,Chacaltana:2011ze,Chacaltana:2012ch,Chacaltana:2013oka,Chacaltana:2014jba,Chacaltana:2014nya,Chacaltana:2015bna}, pioneered by Chacaltana and Distler, describing these theories in detail.

Somewhat unfortunately, these properties of the $T_N$ theory and its cousins were found gradually in the last several years using diverse techniques in various papers.   
The aim of this review article is to collect the most important of these properties, and give a short derivation for each of them from a uniform perspective. The author hopes that a person who would like to join the study of the $T_N$ theories can find this article an easy point of entry. 

The discussions in this article will be based on the following fact:\begin{fact}
The 6d \Nequals{(2,0)} theory of type $G=A_n$, $D_n$ or $E_{6,7,8}$ on $S^1$ is the 5d \Nequals2 supersymmetric Yang-Mills with gauge group $G$.  We write this fact as an equation between quantum field theories: \begin{equation}
S_G\langle S^1\rangle = \SYM_\text{5d \Nequals2}(G).
\end{equation}
Here $S_G$ stands for the 6d \Nequals{(2,0)} theory of type $G$, and the bracket $\langle S^1\rangle$ denotes that the theory is compactified on $S^1$. 
\end{fact} 
In the rest of the article, important results will be summarized similarly as Facts and given in \emph{italic}.  
The derivation for each of the facts might not be quite complete and some of the facts presented might be better termed conjectures.  Some of the \emph{facts} are thus marked with question marks. 
It would be a great pleasure for the author if some of the readers got interested and establish these facts more rigorously.

The rest of the article is organized as follows: 
In Sec.~\ref{sec:construction}, we give a construction of the $T_N$ theory in terms of the 6d \Nequals{(2,0)} theory. 
Namely, we compactify the 6d theory on a Riemann surface of genus $g$ without any punctures. 
We then split them into $2(g-1)$ copies of the $T_N$ theory, corresponding to $2(g-1)$ three-punctured spheres,
and $3(g-1)$ copies of \Nequals2 vector multiplet with gauge group $\SU(N)$. 
We then introduce the concept of the partial closure of punctures, 
and we detail the structure of the associated Nambu-Goldstone (NG) multiplets. 
We conclude the section by a discussion of the Argyres-Seiberg duality.

In Sec.~\ref{sec:centralcharges},
we start from the known anomaly polynomials of the 6d \Nequals{(2,0)} theory to obtain the flavor and conformal central charges of the $T_N$ theory and its partially-closed cousins. 
The discussion in this section improves upon previous discussions in the literature,  by giving a logical derivation of the formulas conjectured  in \cite{Chacaltana:2012zy}. 

In Sec.~\ref{sec:SCI},
we discuss the superconformal indices of the $T_N$ theory and its cousins.
By focusing on the so-called Schur limit, we give a rough derivation of its equivalence with the 2d $q$-deformed Yang-Mills.
The technique is the same: we first consider the case corresponding to a genus $g$ surface without any punctures, 
and then we split them into contributions from copies of the $T_N$ theory and from the vector multiplets.

In Sec.~\ref{sec:moduli}
we study the dimension of the moduli space of supersymmetric vacua of the $T_N$ theory and its cousins,
again by starting from the case corresponding to a surface without any punctures. 
We give an explicit formula for the dimensions of the Higgs branch and the Coulomb branch.
We then discuss the chiral ring relations of the operators on the Higgs branch. 
As our knowledge of these relations is not yet complete, the discussion here will be more schematic than other parts. 
We finish this section by discussion the Seiberg-Witten curves of the partially closed theories. 

In Sec.~\ref{sec:conclusions}
we conclude by listing  papers that describe the properties of the $T_N$ theory not described in this article,
and by giving a short discussion on the future directions of research. 

Before proceeding, we pause here to mention that the statements in the first subsection of each section are applicable to all 4d \Nequals{2} theories in general, whereas the other subsections are mainly for the class S theories. 
We should also mention here that there are other reviews \cite{Tachikawa:2013kta,Gaiotto:2014bja,Neitzke:2014cja,Rastelli:2014jja} on the subjects surrounding the $T_N$ theory, and this article have some overlaps with them. 

\section{The $T_N$ theory and its cousins}\label{sec:construction}

\subsection{Generalities on 4d \Nequals2 theories}
In this review, we often denote a quantum field theory by a letter such as $\cQ$, 
and we sometimes add curly brackets containing the flavor symmetries of the theory: $\cQ\{G\}$ would be a theory with a flavor symmetry $G$.
Unless otherwise mentioned, all quantum field theories we use are 4d \Nequals2  supersymmetric. 
Mostly we only discuss \Nequals2 superconformal theories, and they automatically have $\SU(2)_R\times \U(1)_R$  symmetry. 
We normalize the $\U(1)_R$ charge of a supercharge to be $\pm1$.

We start with a universal fact that is used repeatedly in this review: \begin{fact}\label{mu}
A 4d \Nequals2 superconformal theory $\cQ\{G\}$  has  dimension-2 scalar operators $\mu^{i=+,0,-}$ in the adjoint of $G$ and in the triplet of $\SU(2)_R$, that are in the bottom component of a supermultiplet containing the conserved current of the flavor symmetry $G$.
  The operator $\mu^+$  is chiral in the language of 4d \Nequals1 supersymmetry, and $\mu^-=(\mu^+)^*$ is an anti-chiral operator. 
\end{fact}
The details can be found e.g.~in \cite{Kinney:2005ej} and references therein. As an example, for a free hypermultiplet of charge $+1$ consisting of chiral superfields $Q$ and $\tilde Q$, $\mu^+=Q\tilde Q$ and $\mu^0=|Q|^2-|\tilde Q|^2$.   The $\mu^{+,0,-}$ are often called the moment map operators, since they are the moment maps of the $G$ action on the Higgs branch under the three symplectic forms inherent in the hyperk\"ahler structure. 

We will also use the following: \begin{fact}\label{exact}
Given a 4d \Nequals2 superconformal theory $\cQ\{G\}$, we can couple it to an \Nequals2 gauge multiplet with gauge group $G$.  We denote the resulting gauged theory by $\cQ\{G\} /_\tau G$, where $\tau$ is the coupling constant defined at some renormalization scale. 
When the total one-loop beta function is zero, the coupling constant $\tau$ is exactly marginal.
In this case we say that $\cQ\{G\}$ can be \emph{conformally gauged}.
\end{fact}
When the theory $\cQ\{G\}$ is a free hypermultiplet, we can prove the equivalence of the vanishing of the one-loop beta function and the exact marginality of the coupling constant as follows. 
The Lagrangian of the gauge theory as an \Nequals1 theory roughly has the form \begin{multline}
\left(c\int d^2 \tau \tr W_\alpha W^\alpha + c.c.\right) + c' \int d^4\theta \frac{1}{g^2} \Phi^\dagger \Phi  \\
+ \left(u\int d^2 Q\Phi \tilde Q + c.c.\right) + u' \int d^4\theta \left( Q^\dagger e^V Q + \tilde Q e^V \tilde Q^\dagger\right) .
\end{multline} Using the standard holomorphy arguments, we can show that $\tau$ is renormalized only at one-loop and the coefficient  in front of $Q\Phi \tilde Q$ is not renormalized at all. 
We assume that the theory has zero one-loop beta function, so $\tau$ is not renormalized at all either. 
Now, the \Nequals2 supersymmetry fixes the ratios $c:c'$ and $u:u'$  and therefore nothing is renormalized.
That was what we wanted to show.  
When $\cQ$ is a strongly-coupled field theory, we can use the method of \cite{Green:2010da} to show this fact.

The contribution to the one-loop beta function from the theory $\cQ\{G\}$ is given by the coefficient of the two-point function of the symmetry current. This is also called the flavor symmetry central charge, and we denote it by $k(\cQ)$.
We normalize it so that the contribution from a hypermultiplet in the adjoint representation to be  $k=4h^\vee(G)$, which is $4N$ for $G=\SU(N)$. 
Since the 4d \Nequals4 super Yang-Mills has zero one-loop beta function, the gauge multiplet has $k=-4h^\vee(G)$. 
Then, any theory $\cQ\{G\}$ with $k=4h^\vee(G)$ can be coupled to the \Nequals2 gauge multiplet with gauge group $G$ to have zero one-loop beta function.
We reiterate this as a fact since it is quite important: \begin{fact}
The theory $\cQ\{G\}$ can be conformally gauged if and only if the flavor central charge $k$ of $\cQ\{G\}$ is  $4h^\vee(G)$.  Similarly, if two theories $\cQ_1\{G\}$ and $\cQ_2\{G\}$ have flavor central charge $k_1$ and $k_2$ such that $k_1+k_2=4h^\vee(G)$, we can conformally gauge the diagonal subgroup of two $G$ flavor symmetries. We can denote the resulting theory by 
\removevspaceafterequation
\begin{equation}
(\cQ_1\{G\} \times  \cQ_2\{G\})/_\tau G_\text{diag} .
\end{equation}
\end{fact}

In general, we can characterize exactly marginal deformations of 4d \Nequals2 superconformal theories as follows:
\begin{fact}\label{fact:marginal}
Exactly marginal deformations of a 4d \Nequals2 superconformal theory are in one-to-one correspondence with dimension-2 scalar operators with $\U(1)_R$ charge 4.
\end{fact}
That the bottom component of the supermultiplet containing a marginal deformation is a dimension-2 scalar chiral operator of with $\U(1)_R$ charge 4 is a simple consequence of the structure of short superconformal representations. 
The converse, that such a chiral operator always leads to an exactly marginal deformation, can be shown by the method of \cite{Green:2010da}.

\subsection{The $T_N$ theory}

We introduce  the $T_N$ theory using the theory $S_{\SU(N)}$, the 6d \Nequals{(2,0)} superconformal field theory of type $\SU(N)$. 
Let us  put this 6d theory on a closed Riemann surface $C_g$ of genus $g$. 
When $g\neq 1$, the curvature of the surface $C_g$ breaks all of the supersymmetry.  
We can preserve some part of the supersymmetry by introducing a background R-symmetry gauge field on $C_g$ that partially cancels the curvature.  
Namely, we decompose the $\SO(5)_R$ symmetry of the \Nequals{(2,0)} theory as \begin{equation}
\SO(2)_R\times \SO(3)_R \subset \SO(5)_R,\label{Rdecomp}
\end{equation} and we set \begin{equation}
A_i ^{\SO(2)_R} = -\omega_i^{\SO(2)} \label{Rcancel}
\end{equation}  where $\omega_i$ is the spin connection of $C$. 
This preserves 4d \Nequals2 supersymmetry. 
Indeed, $\SO(2)_R$ and $\SO(3)_R$ of \eqref{Rdecomp} can naturally be identified with $\U(1)_R$ and $\SU(2)_R$ symmetry of the 4d \Nequals2 theory.  
Finally, to isolate a genuine 4d theory, we take the limit where the area of $C_g$ is zero.  
The shape or equivalently the complex structure of $C$ remains to be a physical parameter of the 4d theory. 
We denote the resulting theory $S_{\SU(N)}[C_g]^\IR$.
Here, the bracket operation $\langle C_g\rangle$ stands for putting the theory on the manifold $C_g$, where we keep our choice of the R-symmetry background  implicit in the notation. 
The final superscript $^\IR$ is a reminder that we need to take the 4d limit by taking the area of $C_g$ to be zero.

\begin{figure}[h]
\[
\inc{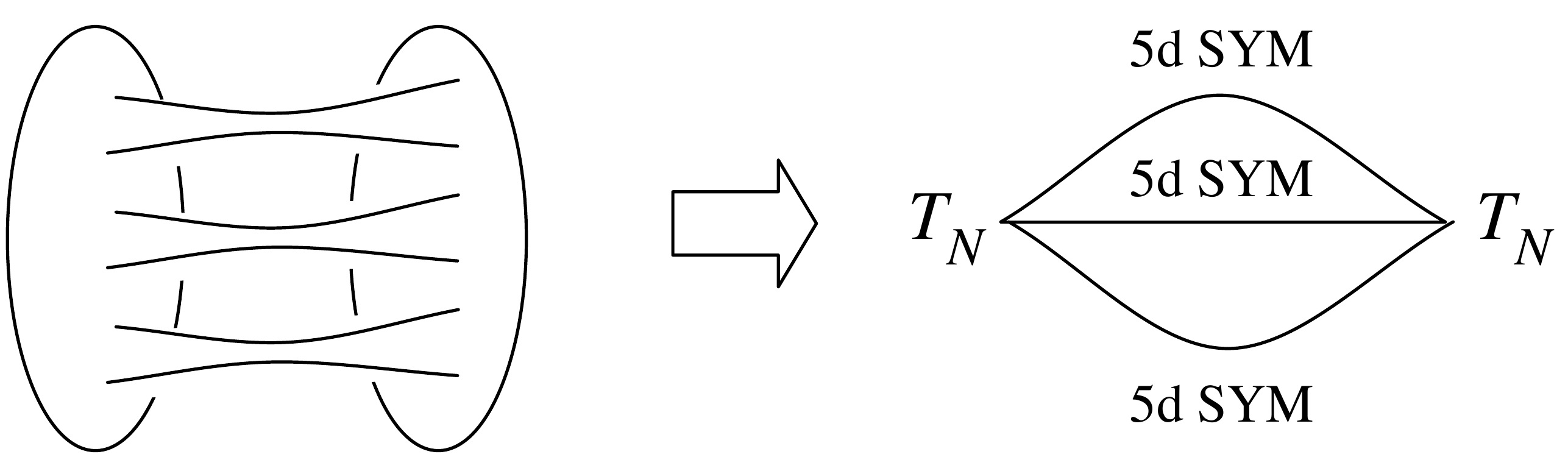}
\]
\caption{6d theory $S_{\SU(N)}$ on a genus-2 surface and its particular limit.\label{fig:alimit}}
\end{figure}

Now we tune the shape of the surface $C$ so that it is composed of $2(g-1)$ three-punctured spheres and  $3(g-1)$  tubes connecting them.  We choose almost all the area of the surface to be in the tubes. 
In this limit, each tube gives a segment of 5d \Nequals2 supersymmetric theory with gauge group $\SU(N)$ with five adjoint scalars $\phi^{i=1,2,3,4,5}$. 
Now, this segment of \Nequals2 super Yang-Mills is  coupled to four-dimensional theories represented by two three-punctured spheres at the two ends, preserving \Nequals2 supersymmetry in 4d.   
There are $3(g-1)$ complex structure deformations of a genus $g$ curve $C_g$,
and in this description they correspond to the length and the twist of the tubes. 
In the 4d language they become $3(g-1)$ exactly marginal deformations.
See  Fig.~\ref{fig:alimit} for a schematic picture for $g=2$. 

We define the $T_N$ theory to be  the 4d limit of the 6d theory on a three-punctured sphere: 
\begin{equation}
T_N = S_{\SU(N)} \langle C_{0,3}\rangle^\IR
\end{equation} where $C_{0,3}$ is the three-punctured sphere. We will introduce other types of punctures later, and this original type of puncture is called  a \emph{full puncture}.

Each tube gives a segment of $\SU(N)$ 5d \Nequals2 theory, and couple two $\SU(N)$ flavor symmetries associated to two punctures. 
In the 4d limit, it reduces to  a 4d \Nequals2 vector multiplet.
To see this, we need to have a better understanding of the coupling of the segment to the 4d \Nequals2 theory at the boundary. 
Such a supersymmetry-preserving boundary condition is roughly described as follows; a similar half-supersymmetric condition of 4d \Nequals4 theory was first discussed in \cite{Gaiotto:2008ak}.

The boundary theory has an $\SU(N)$ flavor symmetry and the bulk $\SU(N)$ gauge field couples to it.  
We split the five scalars $\phi^{i=1,2,3,4,5}$ of the 5d \Nequals2 vector multiplet according to \eqref{Rdecomp} into a doublet  $\phi^{a=1,2}$ of $\SO(2)_R$ and the  triplet $\phi^{i=1,2,3}$ of $\SO(3)_R$.  
Then we put a Neumann boundary condition for $\phi^{a=1,2}$ and a modified version of  Dirichlet boundary condition  for $\phi^{i=1,2,3}$: \begin{equation}
D_n\phi^{a=1,2}|_\text{boundary}=0, \qquad
\phi^{i=1,2,3}|_\text{boundary} = \mu^{i=1,2,3}.\label{5dBC}
\end{equation}  Here, 
the  scalar operators $\mu^{i=1,2,3}$ are the $\SU(2)_R$-triplet scalars associated to the $\SU(N)$ flavor symmetry at the puncture, introduced in Fact~\ref{mu}.

Now, suppose that a tube originally had a radius $R_{S^1}$ and a length $L_\text{segment}$. 
First  reducing it along the $S^1$, we have the 5d \Nequals2 super Yang-Mills with gauge group $\SU(N)$ on a segment, with 5d gauge coupling $1/g_{d=5}^2 \sim 1/R_{S^1}$.  
We now take  the limit where the length $L_\text{segment}$ of the segment is zero. 
We do this in a way that the 4d coupling  $1/g_{d=4}^2 \sim L_\text{segment}/R_{S^1}$ is kept fixed. 
The three scalars $\phi^{i=1,2,3}$ are eliminated due to the Dirichlet boundary condition, 
and the two scalars $\phi^{a=1,2}$ together with the gauge field give rise to the 4d \Nequals2 vector multiplet. 

Summarizing, we see that the theory $S_{\SU(N)}\langle C_2\rangle^\IR$ has a description as two copies of the $T_N$ theory coupled by three $\SU(N)$ 4d \Nequals2 gauge multiplets: \begin{equation}
S_{\SU(N)}\langle C_2\rangle^\IR = (T_N\{G_A,G_B,G_C\} \times T_N\{G_A,G_B,G_C\} ) /_{\tau_A,\tau_B,\tau_C} (G_A \times G_B\times G_C),
\end{equation} where $\tau_{A,B,C}$ are the complexified 4d gauge coupling constants associated to three tubes. 
 
Now we see that the 4d \Nequals2 $\SU(N)$ vector multiplet coupling two $T_N$ theories via two punctures has the gauge coupling constant $1/g_{d=4}^2$ as a tunable parameter. 
This means that the contribution to the one-loop beta function of a puncture is one half of that of an adjoint hypermultiplet.   
Summarizing, we have:\begin{fact}\label{TnDef}
The $T_N$ theory is an 4d \Nequals2 superconformal theory obtained by putting the 6d \Nequals{(2,0)} theory on a three-punctured sphere, with at least $\SU(N)^3$ flavor symmetry and taking the infrared limit: \begin{equation}
T_N:=S_{\SU(N)}\langle C_{0,3}\rangle^\IR.
\end{equation}
  Each puncture carries an $\SU(N)$ flavor symmetry, with the flavor symmetry central charge $k=2N$.
\end{fact}
The $T_N$ theory does not have any exactly marginal deformations, or equivalently, it is an isolated superconformal theory. 
To show this, we just need to show that it does not have a chiral scalar operator with $\U(1)_R$ charge 4.
This is a direct consequence of Fact~\ref{TnCoulomb}, which we will discuss later.

  We similarly define \begin{equation}
  T_G:=S_{G}\langle C_{0,3}\rangle^\IR
\end{equation} for $G=A_{N-1}$, $D_N$, $E_{6,7,8}$. In particular, $T_N=T_{\SU(N)}$.
Most of the discussions below apply equally well to $T_G$ theories for general $G$,
but we often just discuss $T_N$ theories for notational brevity.

In the IR limit we used to define the $T_N$ theory and the $T_G$ theory, the $\SU(2)_R$ that remained unbroken by the background R-symmetry gauge field becomes the $\SU(2)_R$ symmetry of the 4d \Nequals2 super\emph{conformal} symmetry. 
Furthermore, this $\SU(2)_R$ symmetry acts as an $\SO(3)_R$  rotating $\phi_{3,4,5}$ of the 5d \Nequals2 theory that appeared in the intermediate theory.

It should be noted, however, that it is not always the case that this unbroken $\SU(2)_R$ in the compactification becomes the $\SU(2)_R$ of the IR superconformal symmetry.
For example, the compactification on $S^2$ without any puncture gives rise to a hyperk\"ahler sigma model with an intrinsic mass scale in the infrared, and does not lead to a nontrivial superconformal theory.
Also, even when a compactification leads to a nontrivial superconformal theory, the $\SU(2)_R$ in the IR can be different from the $\SU(2)_R$  we just identified in the UV.
For example, the compactification on $T^2$ without any puncture leads to the 4d \Nequals4 theory in the IR, but if viewed as an \Nequals2 theory in the standard manner, the $\SU(2)_R$ in the IR is the subgroup of $\SU(2)_R\times \SU(2)_L \simeq \SO(4)_R$ rotating $\phi_{2,3,4,5}$ of the 5d \Nequals2 theory.

\begin{figure}[h]
\[
\inc{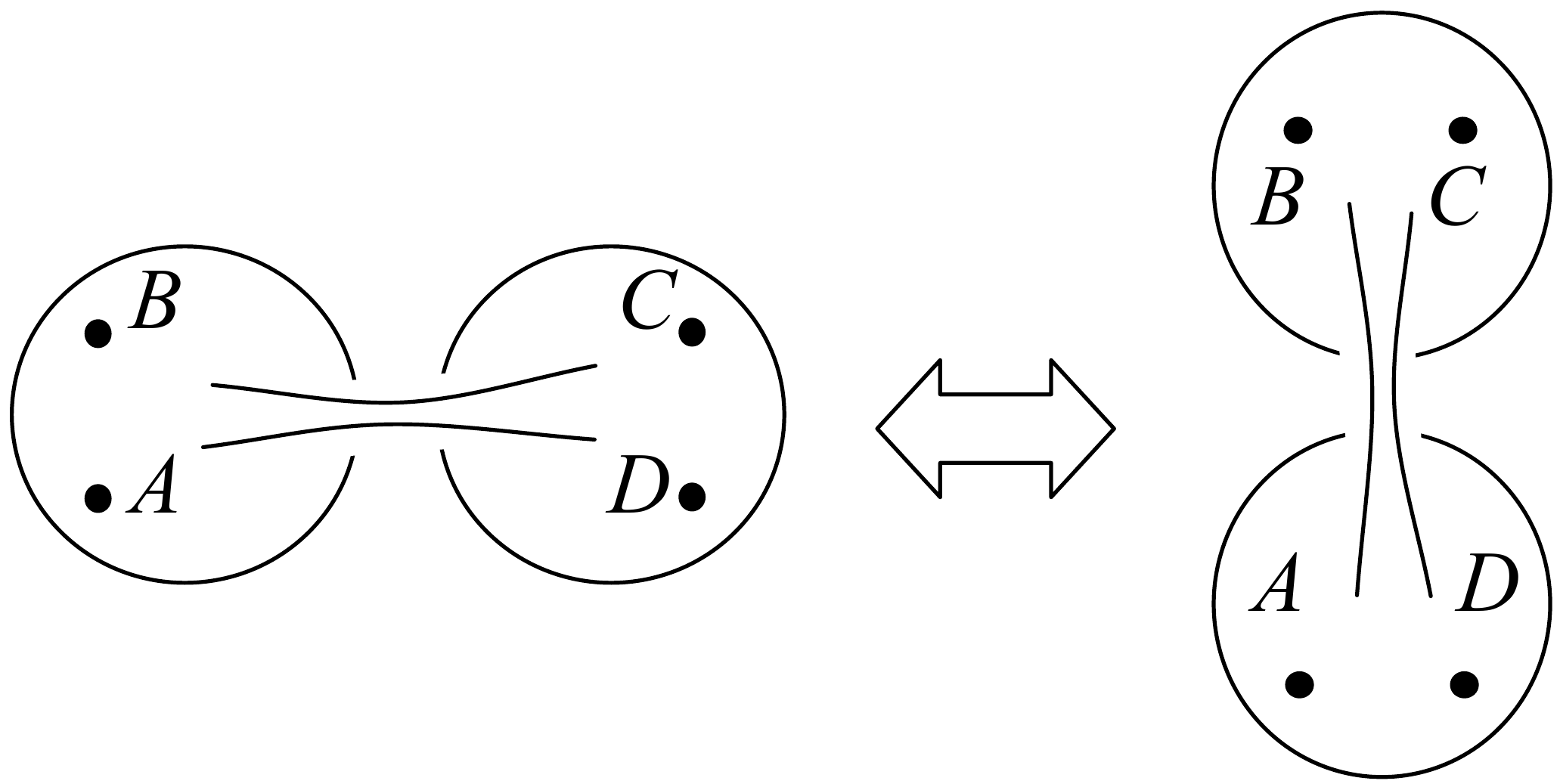}
\]
\caption{Two ways of decomposing a four-punctured sphere.\label{fig:sdual}}
\end{figure}

Now, consider 6d \Nequals{(2,0)} theory of type $\SU(N)$ on a sphere with four full punctures.
Let us use a complex variable $z$ to parametrize the sphere, and put the puncture $A$, $B$, $C$ and $D$ at $z=0$, $z=q$, $z=1$ and $z=\infty$ respectively.  
When $q$ is very small, the theory is given by taking a $T_N\{A,B, G\}$  and another $T_N\{G, C, D\}$ symmetry and by coupling them via an $G_\text{diag}=\SU(N)$ gauge multiplet with the exponentiated coupling constant $q\sim \exp(-1/g_{d=4}^2)$. 
Here we use an abbreviation where $\SU(N)_A$ is written  just as $A$, etc. 
Now, adiabatically change $q$ to be close to $1$; we can now perform the change of coordinates $z'=1-z$  
so that the punctures $A,B,C,D$ are now at $z'=1$, $=1-q$, $=0$ and $=\infty$ respectively. 
Now the theory is given by taking a $T_N\{A,D, G'\}$ and another $T_N\{G', B, C\}$  and by coupling them via an $G'_\text{diag}=\SU(N)_{G'}$ gauge multiplet with the exponentiated coupling constant $1-q\sim \exp(-1/g'_{d=4}{}^2)$. This is a strong-weak duality, or equivalently an S-duality.
Rather than stating this in a sentence, let us write it as an equation:
\begin{fact}\label{fac:sdual}
We have an S-duality  \begin{multline}
(T_N\{A,B,G\} \times T_N\{G,C,D\})
/_q G_\text{diag} \\
= 
(T_N\{A,D,{G'}\} \times T_N\{{G'},B,C\})
/_{1-q} {G'}_\text{diag}.
\end{multline} where $A,B,C,D$ and $G,G'$ are all $\SU(N)$.
\end{fact}

Before proceeding, let us state what $T_2$ and $T_3$ are. We will have more support for these statements later in this review. \begin{fact}\label{t2}
The $T_2$ theory is a theory of four \Nequals2 hypermultiplets. 
In the \Nequals1 language, it consists of eight chiral multiplets $Q_{aiu}$, $a,i,u=1,2,3$ with $\SU(2)^3$ flavor symmetry. 
\end{fact}
\begin{fact}\label{t3}
The $T_3$ theory has an enhanced symmetry  $\SU(3)^3\subset E_6$, and  is the $E_6$-symmetric theory of Minahan and Nemeschansky, originally found in \cite{Minahan:1996fg}. 
\end{fact}

\subsection{Partial closure of punctures}
Let us consider a general situation again: take a 4d \Nequals2 superconformal theory $\cQ\{\SU(N)\}$ with flavor symmetry $\SU(N)$. 
This has a chiral operator $\mu^+$ in the adjoint of $\SU(N)$.  
We are going to give a nilpotent vev to $\mu^+$. 
A nilpotent matrix can be put into the Jordan normal form  \begin{equation}
\vev{\mu^+}=J_Y := J_{n_1} \oplus J_{n_2} \oplus \cdots .
\end{equation} 
where  $N=\sum n_i$, $J_{n}$ is an $n\times n$ Jordan block
with zeros along the diagonal and $n-1$ non-zero entries on one line above the diagonal.
We use $Y$ to denote $n_i$ collectively. 
We can and do order $n_i$ so that $n_1 \ge n_2 \ge \cdots$ without sacrificing generality. 
It is customary to identify $Y$ with a Young diagram such that the $i$-th column has height $n_i$.  It is also customary to abbreviate e.g. the partition $8=3+2+2+1$ as $Y=[32^21]$.
The  Young diagram $Y^t$ transpose to $Y$ is defined by exchanging the rows and the columns. 
Again as an example, $Y^t=[431]$ if $Y=[32^21]$.

Let us first note that when $Y=[1^N]$, it is clearly a trivial operation. This is because  $\vev{\mu^+}=0$, so we do not do anything. Otherwise this is a nontrivial operation. 
The original flavor symmetry $\SU(N)$ is broken to a subgroup $G_Y$, and 
there are Nambu-Goldstone modes and their superpartners associated to this breaking of the flavor symmetry.
 The subgroup $G_Y$ is given by 
 \begin{equation}
G_Y= \mathrm{S}[ \prod_n \U(k_n) ] 
\end{equation} where $k_n$ is the number of times $n$ appears in the sequence $[n_1 n_2\cdots]$.
Here, $\U(k_n)$ acts by permuting the blocks 
\begin{equation}
\underbrace{J_n \oplus  \cdots \oplus J_n}_{k_n}.
\end{equation}
For example, for $N=9$ and $Y=[3^21^3]$, $G_Y=\mathrm{S}[\U(2)\times\U(3)]$. 
We will detail the structure of the Nambu-Goldstone multiplets in Sec.~\ref{sec:NG}.

It turns out to be useful to regard \begin{equation}
J_Y = \rho_Y (\sigma^+)
\end{equation} where $\sigma^+$ is the raising operator of $\SU(2)$ and $\rho_Y:\SU(2)\to \SU(N)$ is an $N$-dimensional representation  of $\SU(2)$, so that we have \begin{equation}
 \mathbf{N}=\bigoplus_i \underline{n_i}.\label{Ndecomp}
\end{equation}  
 Here and below, $\underline{n}$ is an $n$-dimensional irreducible representation of $\SU(2)$.  

As $\vev{\mu^+}=J_Y$ is the highest weight of the $\SU(2)_R$ triplet and the highest weight of $\rho_Y(\SU(2))$ at the same time, the linear combination 
\begin{equation}
I^3 - \frac12\rho_Y(\sigma^3) \label{U1R'}
\end{equation}
of the Cartan part $I_3$ of the $\SU(2)_R$ symmetry  and a Cartan part of the flavor symmetry $\SU(N)$ remains unbroken.
The importance of this unbroken R-symmetry was pointed out in \cite{Gaiotto:2012xa} for example. 
In total, we have the breaking pattern \begin{equation}
\U(1)_R\times \SU(2)_R \times \SU(N) \to \U(1)_R \times \U(1)_R' \times G_Y
\end{equation} where the generator of $\U(1)_R'$ is \eqref{U1R'}.
Note that the chiral supercharges $Q_\alpha^{i=1,2}$ have the charge $(1,\pm1/2)$ under $\U(1)_R\times \U(1)_R'$. 

When the original theory $\cQ\{\SU(N)_1,\ldots,\SU(N)_m\}$ has $\SU(N)^m$ symmetry, we can perform this operation for each $\SU(N)_i$, $i=1,\ldots,m$, by setting $\vev{\mu^+_i}=J_{Y_i}$.  Let us denote the field theory governing everything \emph{except} the NG modes by $\cQ\{Y_1,\ldots,Y_m\}$:
\begin{equation}
\cQ\{\SU(N)_1,\ldots,\SU(N)_m\}
 \xrightarrow{\text{Set $\vev{\mu^+_i}=J_{Y_i}$}}
\cQ\{Y_1,\ldots,Y_m\} + \sum_i \text{(NG modes for $Y_i$)}.
\end{equation} 
At this point, this theory $\cQ\{Y_1,\cdots, Y_m\}$  is a theory with mass scale set by the vev and with the symmetry $\U(1)_R\times \U(1)_R'\times\prod_i G_{Y_i}$; now the generator of $\U(1)_R$ is \begin{equation}
I^3 - \frac12\sum_i \rho_{Y_i}(\sigma^3).
\end{equation}
We are interested mainly in the conformal theories, so let us take the infrared limit of this theory
\begin{equation}
\cQ\{Y_1,\ldots,Y_m\} \xrightarrow{\text{take the IR limit}} \cQ\{Y_1,\ldots,Y_m\}^\IR.
\end{equation}
The resulting theory $\cQ\{Y_1,\ldots,Y_m\}^\IR$ is by definition an \Nequals2 superconformal theory, but it can be free or empty in some special cases.
The procedure of obtaining the new superconformal theory $\cQ\{Y_1,\ldots,Y_m\}^\IR$ from the theory $\cQ$ is called \emph{the partial closure of punctures}. 
Note that this operation, the partial closure of punctures, has mostly been applied only to class S theories in the literature so far, but it can in fact be performed on any 4d \Nequals2 theories. 

In favorable cases, this $\U(1)_R'$ symmetry is the Cartan subgroup of the $\SU(2)_R'$ symmetry of the low-energy \Nequals2 superconformal theory.\footnote{It can happen that $\U(1)_R'$ enhances to $\SU(2)_R'$ only in the infrared limit. It can also happen that $\U(1)_R'$ is already the Cartan of an $\SU(2)_R'$ symmetry before taking the infrared limit.}
Such partial closures are called \emph{good}. 
Otherwise they are called \emph{not good}.\footnote{ Note that even when $\U(1)_R'$ is a part of an  $\SU(2)_R'$ symmery in the ultraviolet, it can happen that this $\SU(2)_R'$ does not survive in the infrared limit.}
When the closures are not good, the low-energy theory $\cQ\{Y_1,\ldots,Y_m\}^\IR$ often has less flavor symmetries than $\prod_i G_{Y_i}$.  

Using the partial closure of punctures, we introduce \begin{fact}
The theories \begin{equation}
T_{Y_1,Y_2,Y_3} := T_N\{Y_1,Y_2,Y_3\}^\IR,
\end{equation}
when good, are \Nequals2 superconformal theories with flavor symmetry at least $G_{Y_1}\times G_{Y_2}\times G_{Y_3}$, obtained by the partial closures of punctures of the $T_N$ theory.
\end{fact}
When $Y_1=Y_2=Y_3=[1^N]$, we do nothing, so we obviously have $T_{Y_1,Y_2,Y_3}=T_N$. 
Another fundamental fact is \begin{fact}\label{fact:bifund}
The theory $T_{[1^N],[1^N],[N-1,1]}$ is a theory of free bifundamental hypermultiplets consisting of \Nequals1 chiral multiplets $Q^i_a$, $\tilde Q^a_i$, $a,i=1,\ldots,N$.
\end{fact}
We will justify this later.  Note that when $N=2$ this fact reduces to Fact~\ref{t2}.

\subsection{Structure of the NG bosons under the partial closures}\label{sec:NG}
Let us study the structure of the Nambu-Goldstone modes that arise associated to the vev $\vev{\mu^+}=J_Y$ by acting them with $\SU(N)$ generators and their superpartners.  
Using the complexified $\SU(N)$ action, i.e.~by using $\SL(N)$ action, 
the vev $J_Y$ can be moved to any  nilpotent matrix conjugate to $J_Y$. 
Let us call the set of all such matrices  the nilpotent orbit $O_Y$ of type $Y$.
From this viewpoint, we picked the vev $ \vev{\mu^+}=J_Y\in O_Y$,
and the Nambu-Goldstone modes correspond to the tangent space at $J_Y$ of $O_Y$.

The directions along the tangent space arise from $\SU(N)$ generators $J^a$ such that \begin{equation}
[\rho_Y(\sigma^+), J^a] \neq 0.
\end{equation} To find them, we just have to decompose the $\SU(N)$ adjoint by regarding it as an $\SU(2)$ representation by $\rho_Y$, and taking non-highest-weight vectors under the $\SU(2)$ action. 

Let us then say that  we have the irreducible decomposition \begin{equation}
\mathbf{adj}=\bigoplus \underline{m_i} \label{adjdecomp}
\end{equation} under $\rho_Y$. 
This decomposition can be easily by plugging \eqref{Ndecomp} to 
$\mathbf{adj}\oplus \bC = \mathbf{N}\otimes \overline{\mathbf{N}}$. 
Each direct summand $\underline{m_i}$ above gives rise to \begin{itemize}
\item $m_i-1$ complex scalars with $\U(1)_R$ charge 0, and $\U(1)_R'$ charge \begin{equation}
\frac{m_i-1}2, \frac{m_i-3}2, \cdots, \frac{3-m_i}2,
\end{equation} 
\item and $m_i-1$ Weyl fermions with $\U(1)_R$ charge $-1$ and $\U(1)_R'$  charge \begin{equation}
\frac{m_i}2-1,\frac{m_i}2-2,\cdots,1-\frac{m_i}2.
\end{equation}
\end{itemize}
The complex dimension of the nilpotent orbit $O_Y$ is then given by the sum $\sum (m_i-1)$. It is a combinatorial exercise to show that \begin{equation}
\dim_\bC O_Y = \sum_i (m_i-1)= N^2-\sum_i s_i^2
\end{equation} where $Y^t=[s_1 s_2 \cdots]$ is the Young diagram transpose to $Y$.
They are free hypermultiplets, but with a slightly unusual assignment of the R-charges. 

Assuming that $\U(1)_R'$ enhances to $\SU(2)_{R}'$ in the infrared, one finds therefore:\begin{fact}\label{fact:NG}
When the partial closure $\vev{\mu^+}=\rho_Y(\sigma^+)$ is good, the resulting Nambu-Goldstone modes 
consist of \begin{itemize}
\item $\U(1)_R$ neutral real scalars in $\underline{m_i-2}\oplus \underline{m_i}$ of $\SU(2)_R'$ and 
\item $\U(1)_R$-charge $-1$ Weyl fermions in $\underline{m_i-1}$ of $\SU(2)_R'$
\end{itemize} for each summand $\underline{m_i}$ in the decomposition  \eqref{adjdecomp}
of the $\SU(N)$ adjoint under $\rho_Y$. In total, there are \begin{equation}
\frac12\dim_\bC O_Y = \frac12(N^2-\sum_i s_i^2)
\end{equation} free hypermultiplets in the Nambu-Goldstone modes. Here $Y^t=[s_1s_2\cdots]$ is the Young diagram transpose to $Y$.
\end{fact}
Note that the description at the first bullet point is not completely precise when $m_i$ is even, since $\underline{m_i}$ and $\underline{m_i-2}$ are not strictly real representations.
Such $\underline{m_i}$'s  appear always in pairs, however, and therefore we have   complex scalars in $\underline{m_i-2}\oplus \underline{m_i}$ for each such pair.

\subsection{Complete closure}
Here let us explain why this operation is called the partial closure. 
Consider the 4d theory $S_{\SU(N)}\langle C_{g,n}\rangle$, obtained by putting the 6d theory 
on a genus $g$ surface with $n$ full punctures.  Then we have the following statement:\begin{fact}\label{fact:empty}
Choose one puncture from the theory $S_{\SU(N)}\langle C_{g,n}\rangle$, 
and perform the closure of type $Y=[N]$ to the $\SU(N)$ symmetry associated to that puncture.  
Then the resulting theory is equivalent to $S_{\SU(N)}\langle C_{g,n-1}\rangle$, where the chosen puncture that was originally full was \emph{completely closed} and disappears:
\removevspaceafterequation
\begin{equation}
S_{\SU(N)}\langle C_{g,n}\rangle\{[N]\} = S_{\SU(N)} \langle C_{g,n-1}\rangle.
\end{equation} 
\end{fact}
At this point we cannot justify this statement except for $N=2$. We will see more justifications later in the review. 

\begin{figure}[h]
\[
\inc{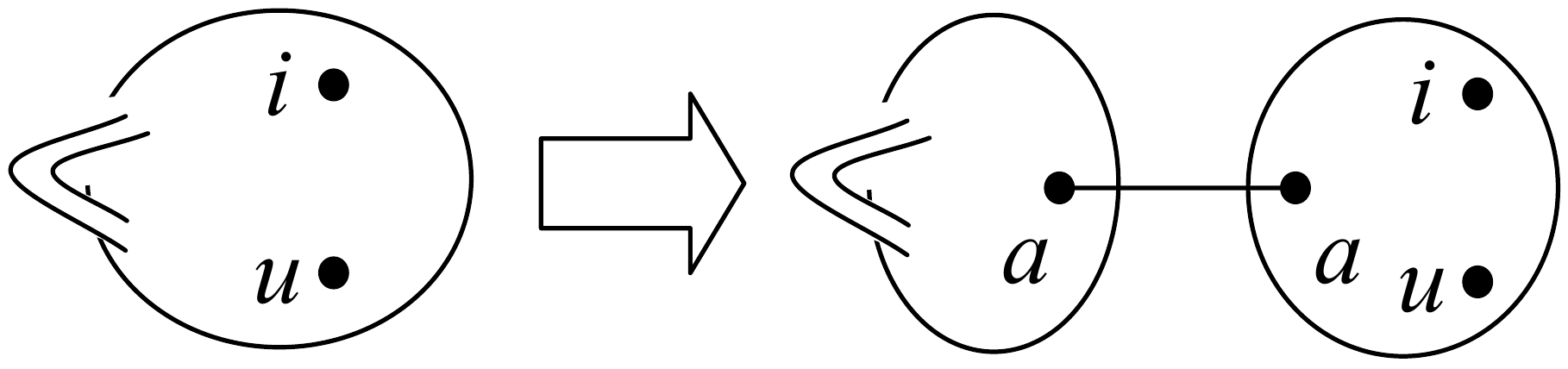}
\]
\caption{Bringing out two full punctures out of a surface.\label{fig:split}}
\end{figure}

So take $N=2$, and assume further that there are $n\ge 2$.  We can modify the shape of the surface so that the theory is given as \begin{equation}
S_{\SU(2)}\langle C_{g,n}\rangle =(  S_{\SU(2)}\langle C_{g,n-1}\rangle \{\SU(2)_a\} 
\times \text{trifundamental $Q_{aiu}$} ) / {\SU(2)_a} \label{closureintermediate},
\end{equation} 
see Fig.~\ref{fig:split}. 
Here we used Fact~\ref{t2} that the $T_2$ theory consists of trifundamental half-hypermultiplets of $\SU(2)^3$.

Now $\mu^+$ associated to the indices $i$ and $u$ are given by \begin{equation}
\mu^+_{(ij)}= Q_{aiu} Q_{bjv} \epsilon^{ab}\epsilon^{uv}, \qquad
\mu^+_{(uv)}= Q_{aiu} Q_{bjv} \epsilon^{ab}\epsilon^{ij}.
\end{equation}  
We now want to close the $\SU(2)_i$ puncture by $Y=[2]$.
Equivalently, ~we would like to set $\vev{\mu^+{}^i_j}=\footnotesize\begin{pmatrix}
0& 1 \\
0&0
\end{pmatrix}$. Since $\mu^+_{ij}=\epsilon_{ik}\mu^+{k}_j$, 
this amounts to setting $\vev{\mu^+_{(11)}} =1$, keeping other components zero.
This can be done by setting $\vev{Q_{aiu}}=\delta_{i=1}\epsilon_{au}$. 

This means that $\SU(2)_a\times \SU(2)_u$  is broken to the diagonal $\SU(2)$ subgroup. So, the $\SU(2)_a$ gauge group is completely Higgsed, eating three hypermultiplets.
Out of the four hypermultiplets in the trifundamental $Q_{aiu}$, only one remains.
Therefore, after setting $\vev{\mu^+}=J_Y$, the theory \eqref{closureintermediate} becomes \begin{equation}
 S_{\SU(2)}\langle C_{g,n-1}\rangle \{\SU(2)_{a=i}\}  + \text{one free hypermultiplet}.
\end{equation} 
The one free hypermultiplet is the Nambu-Goldstone modes associated to the closure by $Y=[2]$. 
We conclude that the resulting theory from the closure is \begin{equation}
S_{\SU(2)}\langle C_{g,n}\rangle\{[2]\} = S_{\SU(2)}\langle C_{g,n-1}\rangle.
\end{equation}

This closure of a puncture of the $T_2$ theory by $Y=[2]$  is our first example of a non-good closure, so let us analyze it more closely.  
We start from the trifundamentals $Q_{aiu}$, and close one puncture by setting $\vev{\mu^+_{(11)}}=1$. 
To separate  the Nambu-Goldstone mode and the rest, note that an infinitesimal complexified $\SU(2)$ action by $a\sigma^+ + b\sigma^3 + c\sigma^-$ changes the vev $\vev{\mu^+}$ by \begin{equation}
\delta\vev{\mu^+_{(11)}} \propto b  , \qquad
\delta\vev{\mu^+_{(12)}} \propto c, \qquad
\delta\vev{\mu^+_{(22)}} \propto 0.
\end{equation} 
Therefore the Nambu-Goldstone mode can be eliminated by requiring \begin{equation}
\vev{\mu^+_{(11)}}=1,\qquad \vev{\mu^+_{(12)}}=0,\label{musep}
\end{equation} keeping $\vev{\mu^+_{(22)}}$ unspecified. 

 In terms of $Z^u_a := \epsilon^{uv} Q_{v,i=1,a}$ 
and $W^u_a :=\epsilon^{uv} Q_{v,i=2,a}$, the equation \eqref{musep} can be written as  \begin{equation}
\det Z=1, \qquad \epsilon_{uv}\epsilon^{ab}Z^{u}_a W^v_b =0.
\end{equation}
The first equation means that $Z$ is on the $\SL(2)=\SU(2)_\bC$ group manifold, and the second equation means that $W$ can be identified as the coordinates of its cotangent bundle. Equivalently, the second equation can be more suggestively written as $d\det Z|_{dZ\to W}=0$, where we first apply the exterior derivative, and then we replace $dZ$ by another commuting variable $W$.

Summarizing, $T_2\{[2]\}$  is an \Nequals2 sigma model on $T^*\SU(2)_\bC$, the cotangent bundle of $\SL(2)=\SU(2)_\bC$. 
In general, we have \begin{fact}
The theory $T_N\{[N]\}$ obtained by the complete closure of a full puncture of the $T_N$  theory is an \Nequals2 sigma model on $T^*\SU(N)_\bC$. It has $N\times N$ chiral fields $Z$ and $W$ satisfying 
\removevspaceafterequation
\begin{equation}
\det Z=1, \qquad d(\det Z)|_{dZ\to W}=0. 
\end{equation}
\end{fact}

The $\SU(N)^2$ symmetry of this theory acts on the fields $Z$ and $W$ naturally by the left and the right actions. But they are not preserved intact at any point: even when $W=0$ and $Z=1$, $\SU(N)^2$ is broken down to the diagonal $\SU(N)$.  To take the low-energy limit, we need to set $\det Z = c^N $, expand $Z=c + \delta Z$,  and send $c\to \infty$, keeping $\delta Z$ and $W$ as the fluctuations.  Only the diagonal $\SU(N)$ is manifest in this limit.

The fact above itself can be shown as follows: 
the $T_N$ theory was for the three punctured sphere, $T_N=S_{\SU(N)}\langle C_{0,3}\rangle^\IR$. 
Therefore, we see that $T_N\{[N]\}=S_{\SU(N)}\langle C_{0,2}\rangle$. 
Now, a sphere with two punctures has an $S^1$ isometry with two full punctures as two fixed points. 
Reducing around this $S^1$, we find that $T_N\{[N]\}$ is essentially the 5d \Nequals2 super Yang-Mills theory with $\SU(N)$ gauge group on a segment.
The boundary condition is in some sense  the opposite of \eqref{5dBC} on both ends:  
we have \begin{equation}
 \phi^{a=1,2}=0, \qquad D_n \phi^{i=1,2,3}=0,
\end{equation} 
and the gauge transformations at the two boundaries are considered as flavor symmetries $\SU(N)\times \SU(N)$. 
We see that the 4d \Nequals2 vector multiplet part is killed by the boundary conditions.
The Higgs branch can be found by studying the moduli space of the BPS equation with these boundary conditions, and turns out to be $T^*\SU(N)_\bC$, see \cite{Gaiotto:2008sa} for more details.

\subsection{Argyres-Seiberg duality}

In Fact~\ref{fac:sdual} we learned an S-duality of two $T_N$ theories coupled by an $\SU(N)$ gauge group.
When $N=2$, this is the standard S-duality of $\SU(2)$ gauge theory with $N_f=4$ flavors, as first beautifully demonstrated in \cite{Gaiotto:2009we}, see also \cite{Argyres:1999fc}. 
However, when $N>2$, this is a duality of non-Lagrangian theories coupled to gauge fields.
Let us now use the partial closure to derive S-dual descriptions of Lagrangian gauge theories.

First, we start from the duality of Fact~\ref{fac:sdual}. 
The basic point was to consider $S_{\SU(N)}\langle C_{0,4}\rangle$, the 6d theory put on a sphere with four punctures $A,B,C,D$, and split the sphere in two ways. 

\begin{figure}[h]
\[
\inc{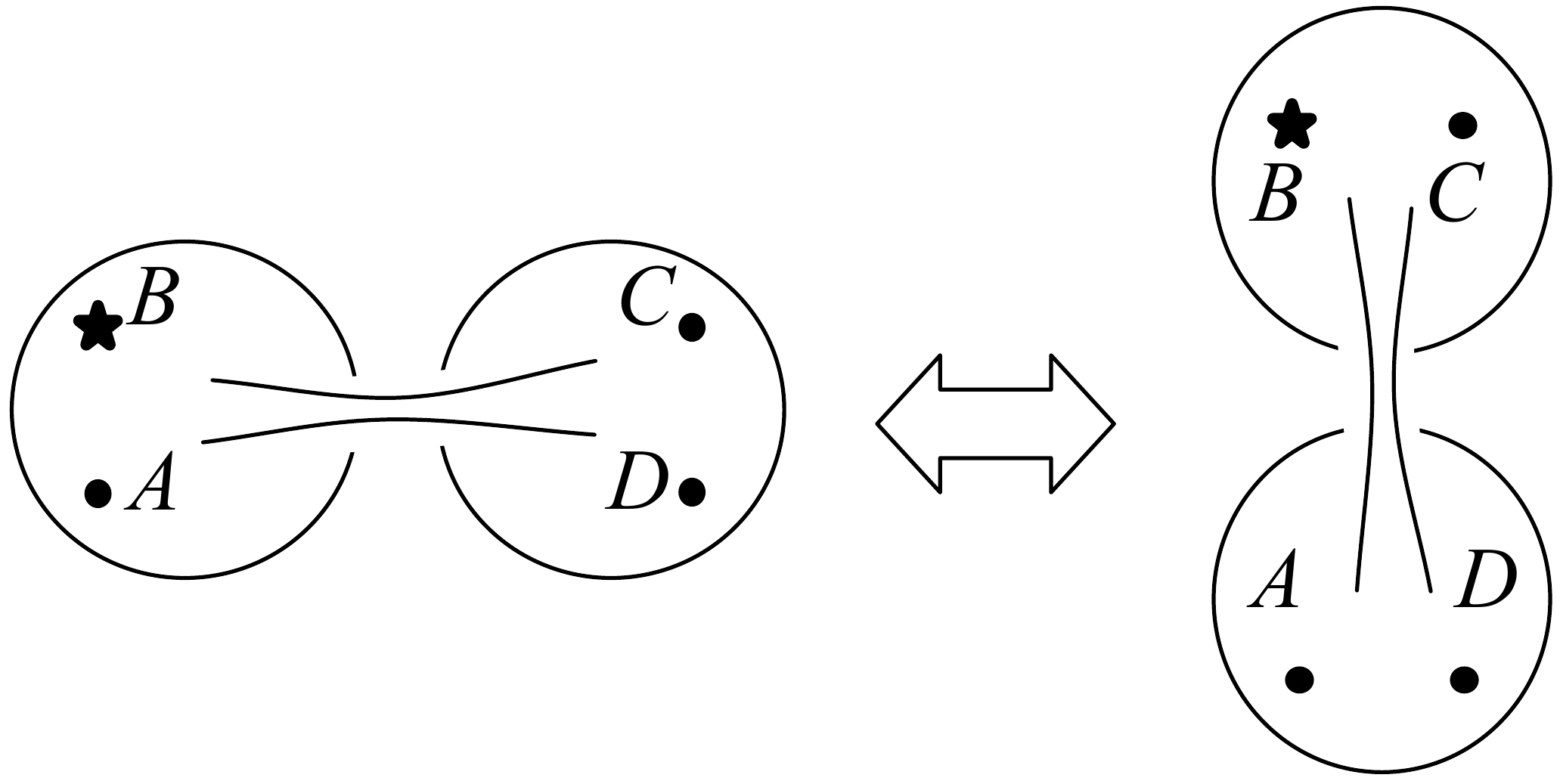}\qquad \inc{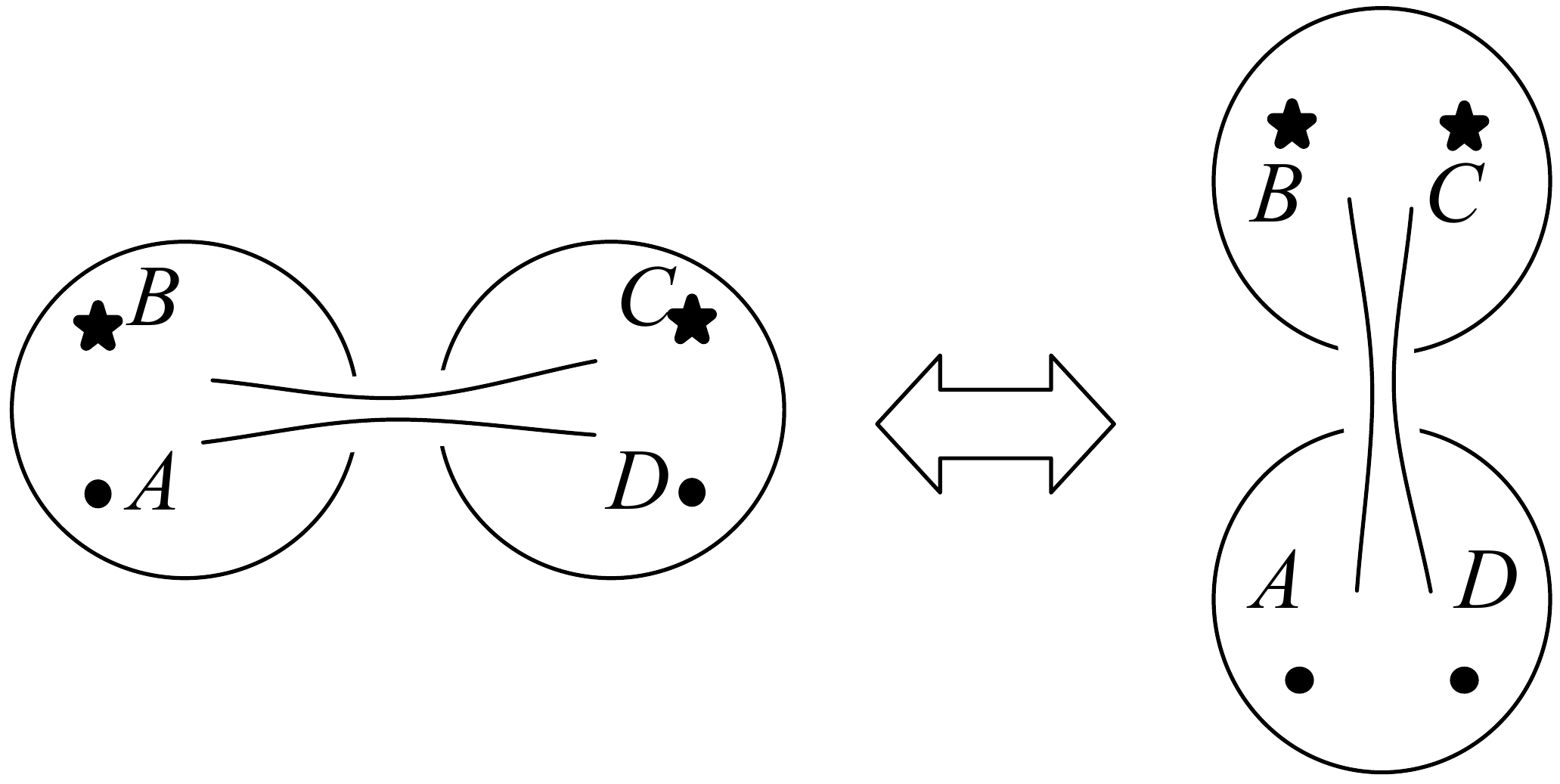}
\]
\caption{We partially close the puncture $B$ to the type $[N-1,1]$.
We then partially close the puncture $C$ to the type $[N-1,1]$. 
The sphere containing two punctures of type $[N-1,1]$ is tricky to analyze.
\label{fig:sdualxy}}
\end{figure}

Now, partially close the puncture $B$ to type $Y=[N-1,1]$, see Fig.~\ref{fig:sdualxy}. 
Considering the splitting in two ways, we have the duality \begin{multline}
(T_N\{A,[N-1,1]_B,G\} \times T_N\{G,C,D\})
/_q G_\text{diag} \\
= 
(T_N\{A,D,{G'}\} \times T_N\{{G'},[N-1,1]_B,C\})
/_{1-q} {G'}_\text{diag},
\end{multline} where $A,D,G,G'$ are $\SU(N)$ gauge or flavor groups.
We stated previously in Fact~\ref{fact:bifund} that  $T_N\{[N-1]\}$ is a theory of bifundamental hypermultiplets of $\SU(N)^2$.
Therefore, this is a duality of the $T_N$ theory coupled to bifundamentals by an $\SU(N)$ gauge group. 

We now further perform the partial closure of the puncture $C$ to $Y=[N-1,1]$.
On the left hand side, we have \begin{equation}
(T_N\{A,[N-1,1]_B,G\} \times T_N\{G,[N-1,1]_C,D\})
/_q G_\text{diag} 
\end{equation} which is just $N_f=N+N$ flavors of fundamental hypermultiplets coupled to $\SU(N)$ gauge group. 
On the right hand side we have \begin{equation}
(T_N\{A,D,{G'}\} \times T_N\{{G'},[N-1,1]_B,[N-1,1]_C\})
/_{1-q} {G'}_\text{diag}
\end{equation} which is more tricky to analyze.
This is because  the partial closure \begin{equation}
T_N\{[N-1,1],[N-1,1]\}
\end{equation} is not good, since $T_N\{[N-1,1]\}$ is already a free theory.
At the same time, because $T_N\{[N-1,1]\}$ is free, we can study this partial closure explicitly.

Denote the bifundamentals as $Q^i_a$ and $\tilde Q^a_i$, 
where the indices $i$, $a$ are  for $\SU(N)_{C}$ and $\SU(N)_{G'}$ respectively. 
We are setting \begin{equation}
Q^i_a \tilde Q^a_j = J_{N-1}\oplus J_1. 
\end{equation}  One way to solve this is to take 
\begin{equation}
Q=\diag(1,1,\ldots,1,0,0), \qquad \tilde Q=J_{N-1}\oplus J_1. 
\end{equation} 
This forces us to have \begin{equation}
\tilde Q^b_i Q^i_a  = J_{N-2} \oplus J_1 \oplus J_1 = J_{[N-2,1^2]}.
\end{equation}  which in turn force $\vev{\mu_+}$ of 
$\SU(N)_G$ of  the first $T_N\{A,D,G'\}$ to be set to \begin{equation}
\vev{\mu_+}  =  J_{[N-2,1^2]}
\end{equation} via the F-term equation of the adjoint scalar of $\SU(N)_{G'}$.
This means that the puncture $G'$ of $T_N\{A,D,G'\}$ is partially closed to type $Y'=[N-2,1^2]$.

Originally, there are $\SU(N)_{G'}\times \SU(N)_C$ symmetry acting on the bifundamental and $\SU(N)_{G'}$ was gauged. 
Now the vev $\vev{Q^a_i}$ and $\vev{\tilde Q^i_a}$ breaks the symmetry down to $\SU(2)\times \U(1)_\text{diag}$ where $\SU(2)\times \U(1)$ is the natural symmetry for $Y'=[N-2,1^2]$ on $\SU(N)_{G'}$,
 $\U(1)$ is the natural symmetry for $Y=[N-1,1]$ on $\SU(N)_C$,
and $\U(1)_\text{diag}$ is the diagonal combination of these two $\U(1)$s. 
In the end, only $\SU(2)_{G'}\subset \SU(N)_{G'}$ remains gauged.

Let us count how many hypermultiplets remain coupled to $\SU(2)_{G'}$. 
Originally we had $x=N^2$ free hypermultiplets in the bifundamental. 
The partial closure of $G'$ to $[N-2,1^2]$ gives $y=N(N-1)/2-3$ Nambu-Goldstone hypermultiplets,
as can be found using Fact~\ref{fact:NG}.
$\SU(N)_{G'}$ is broken to $\SU(2)_{G'}$ and eats $z=N^2-2^2$ hypermultiplets. 
Therefore $x+y-z=N(N-1)/2+1$ free hypermultiplets remain.
Finally the partial closure of $C$ to $[N-1,1]$ gives $w=N(N-1)/2-1$ Nambu-Goldstone hypermultiplets,
so only $x+y-z-w=2$ hypermultiplets remain coupled to $\SU(2)_{G'}$. This is a doublet of $\SU(2)_G$.

Summarizing, we found that coupling $T_N\{G',[N-1,1]_B,[N-1,1]_C\}$ to an $\SU(N)_{G'}$ flavor symmetry via an $\SU(N)_{G'}$ gauge multiplet has the effect that 
the $\SU(N)_{G'}$ is spontaneously broken to $\SU(2)_{G'}$, 
and there is a doublet hypermultiplet coupled to this unbroken $\SU(2)_{G'}$.  Therefore we see that
\begin{fact}\label{fact:AS}
We have an S-duality of the form \begin{equation}
(\text{$2N\times N$ bifundamentals}) /_q \SU(N)
= {T_{[1^N],[1^N],[N-2,1^2]} \times \text{(a doublet)} }/_{1-q} {\SU(2)}
\end{equation}
where the $\SU(2)$ gauge group on the right hand side couples to the $\SU(2)$ flavor symmetry of $[N-2,1^2]$ and to the $\SU(2)$ of the doublet. 
\end{fact} 
For $N=3$, the right hand side slightly simplifies, since $[N-2,1^2]=[1^N]$ in that case.
This the Argyres-Seiberg duality \cite{Argyres:2007cn}. 
For $N\ge 4$ this was described in detail in \cite{Chacaltana:2010ks}.

Note that on the left hand side, there is an $\SU(2N)$ flavor symmetry, while only $\SU(N)\times \SU(N)$ is manifest on the right hand side. 
This means that the $\SU(N)^2$ flavor symmetry of $T_{[1^N],[1^N],[N-2,1^2]}$ should enhance to $\SU(2N)$.  
In particular, for $T_3$, we see that $T_3$ should be such that for any pair of two $\SU(3)$s it should enhance to $\SU(6)$.  
This is only possible when $T_3$ has an $E_6$ flavor symmetry, and supports Fact~\ref{t3}.

Generalizing, it is common that when the punctures $Y_{1,2}$ are rather small, 
the theory $T_N\{\SU(N)_A,Y_1,Y_2\}$ has the effect 
that the first $\SU(N)_A$ is spontaneously broken to a subgroup $H$
and $\vev{\mu^+_A}$ automatically is set to $J_{Y_3}$.
To analyze the theory $S_{\SU(N)}\langle C_{g,n} \rangle \{Y_1,Y_2\}$, 
by taking out the two punctures as in Fig.~\ref{fig:split}, we have \begin{multline}
S_{\SU(N)}\langle C_{g,n} \rangle \{Y_1,Y_2\}\\
=(S_{\SU(N)}\langle C_{g,n-1}\rangle \{\SU(N)_{A'}\} \times T_N\{\SU(N)_A,Y_1,Y_2\})/_q \SU(N)_{A=A'}.
\end{multline}
Now $\vev{\mu^+_A}=J_{Y_3}$ of the second factor causes $\vev{\mu^+_{A'}}=J_{Y_3}$, 
partially closing the first factor and spontaneously breaking $\SU(N)_{A=A'}$ to some subgroup $H_{Y_1,Y_2}\subset G_{Y_3}$.
We end up with the gauge theory \begin{equation}
S_{\SU(N)}\langle C_{g,n} \rangle \{Y_1,Y_2\}
=(S_{\SU(N)}\langle C_{g,n-1}\rangle \{Y_3\} \times \cQ(Y_1,Y_2) )/_q H_{Y_1,Y_2}.
\end{equation} where $\cQ(Y_1,Y_2)$ is a theory determined by $Y_1,Y_2$ with a flavor symmetry $H_{Y_1,Y_2}$.

This point of view was explained e.g.~in \cite{Gaiotto:2011xs}. 
To see which pair of $Y_{1,2}$ leads to this phenomenon of the propagation of partial closure
and what is the resulting $Y_3$, $H_{Y_1,Y_2}$ and the remaining matter theory $\cQ(Y_1,Y_2)$, 
the extensive set of tables in \cite{Chacaltana:2014jba,Chacaltana:2013oka,Chacaltana:2012ch,Chacaltana:2011ze,Chacaltana:2010ks,Chacaltana:2015bna}
is very useful. 
Note however that in these papers, such pair $Y_{1,2}$ is said to require an irregular puncture $Y_3^*$ dual to $Y_3$, 
$\cQ(Y_1,Y_2)$ is listed as a theory of the three-punctured sphere with punctures of type $Y_1$, $Y_2$, $Y_3^*$,
and the group $H_{Y_1,Y_2}$ is listed as a cylinder connecting $Y_3$ and $Y_3^*$.
Note also that the irregular punctures in their terminology are not the same concept as the irregular punctures 
as used e.g.~in \cite{Gaiotto:2009hg,Gaiotto:2009ma,Bonelli:2011aa,Xie:2012hs}.

\section{Central charges}\label{sec:centralcharges}
\subsection{Generalities on the central charges and anomalies}
Four-dimensional conformal theories have two conformal central charges $a$ and $c$.  
For \Nequals2 superconformal theory, using $n_h$ and $n_v$ are more convenient, normalized so that $(n_h,n_v)=(1,0)$ for a free hypermultiplet and $=(0,1)$ for a free vector multiplet;   $a$ and $c$  can be written as \begin{equation}
a=\frac1{24}n_h+\frac5{24}n_v\qquad
c=\frac1{12}n_h+\frac16n_v.
\end{equation} 
The current two-point function of a flavor symmetry $G$ is also characterized by a number $k$, called the flavor central charge. 
As already stated, we normalize it so that  $k=4h^\vee(G)$ for an adjoint hypermultiplet.

In this review, instead of computing the central charges directly, we use the following relations of the central charges and the 't Hooft anomaly of the theory:\begin{fact}
The anomaly polynomial $A_6$ of 4d \Nequals2 theory $\cQ\{G\}$  with flavor symmetry $G$ 
has the following form \begin{equation}
A_6=(n_v-n_h) (\frac{c_1^3}3-\frac{c_1}{12}p_1) - n_v c_1 c_2 +  k c_1 n(F_G) .
\end{equation}
Here,  $c_1=c_1(F_{\U(1)R})$, $c_2=c_2(F_{\SU(2)R})$ are the Chern classes of the background $\U(1)_R$ and $\SU(2)_R$ gauge fields  and $p_1=p_1(TX)$ is the Pontrjagin class of the spacetime,
and $n(F_G)$ is the characteristic class for the background flavor symmetry gauge field 
proportional to $\tr F_{G}^2$ so that it integrates to 1 in the one-instanton background. 
In particular, we have $n(F_G)=c_2(F_G)$ when $G=\SU(2)$.
\end{fact}
The essential analysis establishing this fact was performed in \cite{Kuzenko:1999pi}.

\subsection{The central charge of the $T_N$ theory}
Let us determine the central charge of the $T_N$ theory. 
As always our strategy is to start from the 6d theory.
We first quote the known anomaly polynomial of the 6d \Nequals{(2,0)} theory of type $G$:\begin{fact}
The 6d \Nequals{(2,0)} theory $S_G$ of type $G=A_{N-1}, D_N, E_{6,7,8}$ has the anomaly polynomial 
\begin{equation}
A_8=\frac{h^\vee d}{24} p_2(NY) + r I_8^\text{free}, 
\end{equation}
Here, $Y$ is the spacetime, $TY$ is the tangent bundle, $NY$ is the $\SO(5)_R$ bundle and
\begin{equation}
I_8^\text{free}=\frac{1}{48}\left(p_2(NY)-p_2(TY)+\frac14(p_1(NY)-p_1(TY))^2\right)
\end{equation}
is the anomaly polynomial of the free \Nequals{(2,0)} tensor multiplet.
As always, $h^\vee$, $d$, and $r$ are the dual Coxeter number, the dimension, and the rank of  $G$,
and $h^\vee=N$, $d=N^2-1$ and $r=N-1$ for $G=\SU(N)$.
\end{fact}
This formula was first found for $G=A_N$ in \cite{Harvey:1998bx} using M-theory, conjectured generally in \cite{Intriligator:2000eq}, and computed for $G=D_N$ in \cite{Yi:2001bz}. A field theoretical derivation was given in \cite{Ohmori:2014kda,Intriligator:2014eaa}.

From this we can easily find the anomaly polynomial in four dimensions \cite{Benini:2009mz,Alday:2009qq}. We assume that the 6d spacetime is of the form $Y_6=X_4\times C_2$. 
Note that in this section the subscript of $C_2$ stands for the dimensionality, not the genus.
We arrange the $\SO(5)_R$ bundle $NY$ to partially cancel the curvature of $C_2$ as specified in \eqref{Rdecomp}, \eqref{Rcancel}. Then we just integrate the resulting $I_8$ over $C_2$, obtaining $A_6$. 

In the actual computation, it is convenient to use the so-called splitting principle used in the algebraic topology when manipulating the characteristic classes. 
This principle says that the computation of the characteristic classes can be done  assuming that the vector bundles are just direct sums of line bundles. 
The curvature of those constituent line bundles are called Chern roots.

Let us denote the Chern roots of $TX_4$, $TC_2$ and $NY_6$ by $\pm \lambda_{1,2}$, $\pm t$, $\pm n_{1,2},0$  respectively. 
We also introduce the $\U(1)_R$ bundle and the $\SU(2)_R$ bundle on $X_4$; let us denote their Chern roots by $c_1$ and $\pm \alpha$ respectively. We need to express $n_{1,2}$ in terms of $c_1$, $\alpha$ and $t$. 
The cancellation of the curvature \eqref{Rcancel} in this language is to take $n_1|_{C_2}=-t$.  
Then we identify the 4d R-symmetries with the subgroup of the 6d R-symmetry via \eqref{Rdecomp}.
Therefore we have \begin{equation}
n_1=2c_1-t, \qquad n_2=2\alpha. 
\end{equation}

The anomaly $A_8$ can now be written in terms of the Chern roots by using the following facts: 
When an $\SO$ bundle $B$ has the Chern roots $\pm \lambda_i$, $p_1(B)=\sum \lambda_i^2$ and $p_2(B)=\sum_{i<j} \lambda_i^2\lambda_j^2$. 
Also, for an $\SU(2)$ bundle $R$ with the Chern roots $\pm \alpha$, $c_2(R)=-\alpha^2$. 

Plugging them into $A_8$ and integrating over $C_2$ using Gauss-Bonnet theorem  $\int_{C_2}  t=2-2g$, we find \begin{equation}
A_6
=(g-1) r (\frac{c_1^3}3-\frac{c_1}{12}p_1)-(g-1)(\frac43h^\vee d+r) c_1 c_2.
\end{equation} 
Therefore, the  theory $S_G\langle C_g\rangle$ has \begin{equation}
n_v=(g-1)(\frac43h^\vee d+r),\qquad
n_h=(g-1)\frac43h^\vee d.
\end{equation}

We know that this theory is composed of $2(g-1)$ copies of the $T_G$ theory and $3(g-3)$ tubes each representing  a vector multiplet of gauge group $G$. 
We already determined the flavor central charge of three $G^3$ flavor symmetries. 
Summarizing, we have
\begin{fact}\label{TGnvnh}
The $T_G$ theory has the central charges \begin{equation}
n_v=\frac23h^\vee d + \frac r2 -\frac 32 d,\qquad n_h=\frac23 h^\vee d, \qquad k_{1,2,3}= 2h^\vee
\end{equation} where $k_{1,2,3}$ are the current algebra central charges for three $G$ symmetries. 
\end{fact}

Take $G=\SU(2)$. We find $n_v=0$ and $n_h=4$.  In general, any \Nequals2 superconformal theory has nonnegative $n_v$ and $n_h$ \cite{Hofman:2008ar,Shapere:2008zf}. 
The converse is a conjecture: 
\begin{fact?}
An \Nequals2 superconformal theory with $n_v=0$ is a theory of free hypermultiplets. 
Similarly, when $n_h=0$, it is a theory of free vector multiplets.   
\end{fact?}
Assuming this, we find that the $T_2$ theory consists of four free hypermultiplets with $\SU(2)^3$ symmetry, such that each of $\SU(2)$ has $k=4$. The trifundamental $Q_{aiu}$ is the only such multiplet, supporting our  Fact~\ref{t2}.

Using the central charges of the $T_G$ theory obtained above, it is easy to get the general formula of the central charges of the theory $S_G\langle C_{g,n}\rangle$. 
In the 4d language, it can be made from $2(g-1)+n$ copies of the $T_G$ theory and $3(g-1)+n$ vector multiplets with gauge group $G$. 
In total, we have \begin{equation}
n_v=(g-1)(\frac43h^\vee d+r)+(\frac23h^\vee d+\frac{r-d}2)n, \quad
n_h=(g-1)\frac43h^\vee d + \frac23h^\vee d n; \label{nvnhfull}
\end{equation}  
The flavor symmetry $G_i$ associated to the $i$-th full puncture has $k_i=2h^\vee$, as always. 

\subsection{Effect of the complete closure}
We would like to know the effect of the partial closures to these central charges, assuming that the closures are good.
Let us first study the effect of the complete closure of a puncture. For definiteness take $G=\SU(N)$.
Originally the contribution to the anomaly polynomial from a full puncture is \begin{equation}
A_6 = \frac{1}2(r-d) (\frac{c_1^3}3-\frac{c_1}{12}p_1) - (\frac23h^\vee d+\frac{r-d}2) c_1 c_2 + 2h^\vee c_1 c_2(\SU(N)), \label{fullcontrib}
\end{equation} as can be seen from \eqref{nvnhfull}.
We set $\vev{\mu^+}=J_N=\rho_{[N]}(\sigma^+)$.
The $\SU(2)_R'$ of the infrared is essentially the diagonal combination of $\SU(2)_R$ before the closure and $\rho_{[N]}(\SU(2))$. 
Stated differently, in terms of the Chern roots $(+\alpha,-\alpha)$ of the infrared $\SU(2)_R'$, 
the Chern roots of the original $\SU(2)_R$ are $(+\alpha,-\alpha)$ and those of $\SU(N)$ are \begin{equation}
(N-1)\alpha, (N-3)\alpha,\ldots, (1-N) \alpha. 
\end{equation} 
We then use that the instanton number $n(F_G)$ of a bundle with Chern roots $\alpha_i$ is \begin{equation}
n(F_G)= -\frac12\sum \alpha_i^2.
\end{equation}
Plugging everything in to \eqref{fullcontrib}, and using $r=N-1$, $h^\vee=N$ and $d=N^2-1$, we find \begin{equation}
A_6=-\frac{N(N-1)}2 (\frac{c_1^3}3-\frac{c_1}{12}p_1) +\frac16(N-1)N(2N^2-2N-1) c_1 c_2.\label{I6tmp}.
\end{equation}
Note that this is the anomaly polynomial contribution from the puncture of type $[N]$ together with the Nambu-Goldstone bosons.

To determine the contribution from the latter,  we note that the decomposition of the adjoint \eqref{adjdecomp} in this case is \begin{equation}
\mathbf{adj}=\bigoplus_{i=2}^N \underline{2i-1}
\end{equation} and therefore the Weyl fermions in the Nambu-Goldstone multiplets have $\U(1)_R$ charge $-1$ and in the $\SU(2)_R'$ representation $\bigoplus_{i=1}^{N-1} \underline{2i}$. 
So the contribution to the anomaly from the Nambu-Goldstone modes are \begin{equation}
A_6=-\sum_{i=1}^{N-1} 2i  (\frac{c_1^3}3-\frac{c_1}{12}p_1) 
+ c_1 c_2 \sum_{i=1}^{N-1} \frac16 (2i-1)2i (2i+1) 
\end{equation} which is precisely equal to \eqref{I6tmp}. 
This means that the contribution to the anomaly from the puncture of type $[N]$, without the Nambu-Goldstone multiplets, is exactly zero.  
This computation supports Fact~\ref{fact:empty} that having the puncture of type $[N]$ is equivalent to having no puncture at all. 

\subsection{Effect of the partial closure to $[N-1,1]$}
Let us next consider the partial closure to $[N-1,1]$. 
We use the embedding $\rho_{[N-1,1]}:\SU(2)\to \SU(N)$. This preserves a $\U(1)_B$ subgroup as the flavor symmetry. 
Now the Chern roots for the $\SU(N)$ flavor symmetry is \begin{equation}
(N-2)\alpha+\beta, (N-4)\alpha+\beta,\ldots, (2-N) \alpha+\beta, (1-N)\beta,
\end{equation} 
where $\pm\alpha$ are the Chern roots for the $\SU(2)_R'$ in the infrared,
and $\beta$ is the Chern root of the $\U(1)$ flavor symmetry. 
Then the total $A_6$ is \begin{equation}
-\frac{N(N-1)}2 (\frac{c_1^3}3-\frac{c_1}{12}p_1) +\frac16(N-1)N(2N^2-8N-1) c_1 c_2 -N^2(N-1)c_1 \beta^2.\label{simptmp1}
\end{equation}
The decomposition of the adjoint under $\SU(2)_R'$ is \begin{equation}
\textrm{adj}=\underline 1\oplus \underline{N-1}\oplus \underline{N-1} \oplus\bigoplus_{i=2}^{N-1} \underline{2i-1}.
\end{equation} 
The Weyl fermions in the Nambu-Goldstone modes therefore have the $\SU(2)_R'$ representations \begin{equation}
\underline{N-2}\oplus \underline{N-2}\oplus \bigoplus_{i=1}^{N-2} \underline{2i},
\end{equation} where two $\underline{N-2}$ terms have $\U(1)_B$ charge $\pm N$ and the other terms are neutral. 
The contribution to the anomaly is then \begin{multline}
A_6=
-\frac{(N+1)(N-2)}2 (\frac{c_1^3}3-\frac{c_1}{12}p_1)\\
+\frac16(N-1)(N-2)(2N^2-4N-3) c_1 c_2
-N^2(N-2) c_1 \beta^2.\label{simptmp2}
\end{multline}
Subtracting \eqref{simptmp2}  from \eqref{simptmp1}, we find that the contribution to the anomaly from the puncture of type $[N-1,1]$ is \begin{equation}
-(\frac{c_1^3}3-\frac{c_1}{12}p_1)+ (1-N^2)c_1 c_2 -N^2 c_1\beta^2.
\end{equation}

From this we can find the central charges of the theory $T_N\{[N-1,1]\}$ by adding three contributions:
\begin{multline}
A_6 = -(N-1)  (\frac{c_1^3}3-\frac{c_1}{12}p_1) + (\frac43 N^3-\frac N3 -1) c_1 c_2 \\
+ 2( - \frac{N^2-N}2 (\frac{c_1^3}3-\frac{c_1}{12}p_1) - (\frac 23 n^3 - \frac{N^2}2-\frac N6) c_1 c_2  ) \\
-(\frac{c_1^3}3-\frac{c_1}{12}p_1)+ (1-N^2)c_1 c_2 -N^2 c_1\beta^2.
\end{multline} Here the first line is the part proportional to $2-2g$, the second line is from two full punctures, and the third line is from the puncture of type $[N-1,1]$. The final answer is rather simple: \begin{equation}
= -N^2  (\frac{c_1^3}3-\frac{c_1}{12}p_1) -N^2 c_1\beta^2. 
\end{equation} 
Therefore this theory has $n_v=0$, $n_h=N^2$. This strongly suggests that the theory consists of $N^2$ free hypermultiplets.
Then the term proportional to $\beta^2$ says that the $\U(1)_B$ charges of the hypermultiplets are $\pm1$.
In addition, two $\SU(N)$ symmetries both have $k=2N$.  
This supports the fact that this theory consists of free hypermultiplets in the bifundamental of $\SU(N)\times \SU(N)$, and the $\U(1)_B$ charge carried by the puncture of type $[N-1,1]$ can be identified with the baryonic symmetry of the bifundamental.  

\subsection{General formula}
From the examples above, it is clear that we can compute the superconformal central charges $n_v$, $n_h$ and the flavor symmetry central charges $k_i$ of the theories with partially closed punctures, by identifying $\SU(2)_R'$ after the closure in the original variables and subtracting the contributions from the Nambu-Goldstone multiplets.  Instead of giving a detailed derivation we just quote the facts:
\begin{fact}
The central charges $n_v$, $n_h$ of the 4d theory obtained by putting the 6d theory of type $G$ on a genus $g$ surface with $n$ punctures, labeled by $Y_1$, \ldots, $Y_N$, are given by 
\begin{equation}
n_v=(g-1)(\frac43h^\vee d+r)+\sum_i n_v(Y_i), \quad
n_h=(g-1)\frac43h^\vee d + \sum_i n_h(Y_i) 
\end{equation} where \begin{equation}
n_v(Y)=\frac23h^\vee d -  4 \rho_W \cdot h_Y + \frac 12 (r-n_o(Y)),
n_h(Y)=\frac23h^\vee d -  4 \rho_W \cdot h_Y + \frac 12 n_e(Y) .\label{nvY}
\end{equation}
Here $\rho_W$ is the Weyl vector of $G$, $h_Y$ is  the highest element in the Weyl orbit of $\rho_Y(\sigma_3)$, $n_{o,e}(Y)$ are the number of direct summands in the decomposition of the adjoint \eqref{adjdecomp} under $\rho_Y$, that are respectively odd and even dimensional. 
When $G=\SU(N)$,  $\rho_W=(N-1,N-3,\ldots,1-N)/2$ and
$h_Y$ is the vector $\rho_Y(\sigma_3)$ reordered so that the components are non-decreasing, e.g. $h_{[3,1]}=(1,0,0,-1)$. 
\end{fact}

This general form of the $n_{v,h}$ was originally derived in \cite{Chacaltana:2012zy} using various string dualities. Here we instead gave a derivation using the Nambu-Goldstone multiplets.\footnote{The author should confess that he has not combinatorially proved that the formula \eqref{nvY} results from the analysis of Nambu-Goldstone multiplets. At least he checked the validity in numerous cases.}
Similarly, we have the following facts concerning the flavor symmetry central charge:\begin{fact}
For a puncture of type $Y=[n_1 n_2 \cdots ]$ and a factor the flavor symmetry $\SU(\ell)$ associated to the $\ell$ columns of height $h$, its flavor symmetry central charge $k_{\SU(\ell)}$ is given by \begin{equation}
k_{\SU(\ell)}= 2 \sum_{ h' \le h } s_{h'}
\end{equation}  where $Y^t=[s_1s_2\cdots]$ is the Young diagram transpose to $Y$. 
\end{fact}

As an example, consider a puncture of type $Y=[N-2,1,1]$. The $\SU(2)$ symmetry associated to two columns of height $1$ then has $k=6$, since $Y^t=[3,1^{N-3}]$.
This is nicely consistent with the Argyres-Seiberg duality we reviewed as Fact~\ref{fact:AS}.
Indeed, in the second line, the $\SU(2)$ gauge group couples to the $\SU(2)$ flavor symmetry of a puncture of type $[N-2,1,1]$ and to a doublet.
The one-loop beta function contribution from the matter sector is therefore $6+2=8$, which means that this combined $\SU(2)$ symmetry can be conformally gauged. 

\section{Superconformal index}\label{sec:SCI}
\subsection{Generalities on the superconformal index}
In this section we summarize the superconformal index of the $T_N$ theory and its cousins. 
The first study of this topic was in \cite{Gadde:2009kb}, and the full structure began to emerge in \cite{Gadde:2011ik}.
The main original reference is \cite{Gadde:2011uv} and a nice review can be found in \cite{Rastelli:2014jja}.\footnote{Note that $I_3=R_\text{there}$ and $r=2r_\text{there}$. 
Also beware that the definitions of $p$, $q$ and $t$ in their series of papers \emph{before} \cite{Gadde:2011uv} fluctuated greatly.} 
We concentrate on the so-called Schur limit, which can be introduced most logically at the technology currently available. 

For a 4d \Nequals2 superconformal theory with flavor symmetry $G_F$, its superconformal index is 
a Witten index with respect to a carefully chosen supercharge. 
Let us define a function on four variables $s$, $p$, $q$, $t$ and $g\in G_F$ by \begin{equation}
I(s,p,q,t;g)=\tr_{H(S^3)} (-1)^F s^{\Delta/2-j_2-I_3+r/4}
p^{\Delta/2+j_1-I_3-r/4}
q^{\Delta/2-j_1-I_3-r/4}
t^{I_3+r/2}
g .\label{pqtSCI}
\end{equation}
Here, $H(S^3)$ is the Hilbert space of the theory on $S^3$, or equivalently the space of operators; 
$j_{1,2}$ are the spins of the spacetime $\SO(4)\simeq \SU(2)_1\times \SU(2)_2$, $I_3$ is the spin under $\SU(2)_R$, $r$ is the $\U(1)_R$ charge normalized so that the supercharges have charge $\pm1$. 
The  exponent of $s$ is $\{Q_{1\dot-},(Q_{1\dot-})^\dagger\}$, and  the exponents of $p$, $q$, $t$ together with $g$ all commute with $Q_{1\dot-}$. 
As such, it is invariant under all the exactly marginal deformations and independent of $s$.  
This defines the superconformal index that depends on three variables $p$, $q$ and $t$.

The superconformal index with three variables is not completely understood, but the particular limit $q=t$ is well-understood. 
Let us set $q=t$ and replace $s$ by $s/q$ in \eqref{pqtSCI}: \begin{equation}
I(s/q,p,q,q;g) =\tr_{H(S^3)} (-1)^F s^{\Delta/2-j_2-I_3+r/4} 
p^{\Delta/2+j_1-I_3-r/4}
q^{\Delta-I_3}
g.\label{schurtmp}
\end{equation}  
The exponent of $p$ is $\{Q_{1+},(Q_{1+})^\dagger\}$, and now both $Q_{1+}$ and $Q_{1\dot -}$  commute with the exponent $j_2-j_1+I_3$ of $q$. 
Therefore, the expression above is automatically independent of both $s$ and $p$.  
This limit is often called the Schur limit.  
Summarizing, 
\begin{fact}
Given a 4d \Nequals2 superconformal theory $\cQ\{G\}$ with symmetry $G$, the superconformal index in the Schur limit is defined by \begin{equation}
I_{\cQ}(g)=\tr_{H_{\cQ}(S^3)}(-1)^F q^{\Delta-I_3} g.
\end{equation}  This is essentially the partition function of the theory $\cQ$ on $S^1\times S^3$, and \begin{equation}
q=e^{-2\pi R_{S^1}/R_{S^3}}.
\end{equation} 
We keep the argument $q$ to the index implicit. 
\end{fact}
Here we set $s=p=q$ in \eqref{schurtmp}, and determined the relation between $q$ and the radii of $S^1$, $S^3$ by the conformal mapping. 
For free hypermultiplets we can easily compute this trace to have the following fact:\begin{fact}\label{fact:scihyp}
A hypermultiplet, containing \Nequals1 chiral multiplets in the representation $R\oplus \bar R$ of a symmetry $G$, has the index \begin{equation}
I(g)=\prod_{w} \prod_{n\ge 0} \frac{1}{1-q^{n+1/2} g^w}
\end{equation} where 
$w$ runs over the weights of $R\oplus \bar R$, 
$g$ is now regarded as a Cartan element $g=(z_1,z_2,\ldots,g_r)$ of the symmetry group $G$, 
and $g^w:=\prod z_i^{w_i}$ where $w=(w_1,\ldots,w_r)$. 
\end{fact}

For example, a trifundamental half-hypermultiplet of $\SU(2)^3$ has the index \begin{equation}
I(a,b,c)=\prod_{\pm\pm\pm}\prod_{n\ge 0} \frac{1}{1-q^{n+1/2}a^\pm b^\pm c^\pm}.
\end{equation}

Take an \Nequals2 superconformal theory $\cQ\{G,H\}$ whose  flavor symmetry $G$ can be conformally gauged. 
Then the gauge theory $\cQ/G\{H\}$ is itself a superconformal theory with flavor symmetry $H$ where the coupling constant is exactly marginal.
The superconformal index of the resulting theory can then be computed in the limit where the vector fields are very weakly coupled. 
The result can be summarized as follows:
\begin{fact}\label{fact:scigauging}
When the theory $\cQ$ has flavor symmetry $\SU(N)\times H$ and the  $\SU(N)$ can be conformally gauged, 
the theory $T/\SU(N)$  with flavor symmetry $H$ has the superconformal index given by \begin{equation}
I_{\cQ/\SU(N)}(h)=\frac{1}{N!} \oint \prod_{i=1}^{N-1} \frac{dz_i}{2\pi \sqrt{-1} z_i}
\prod_{i\neq j} (1-\frac{z_i}{z_j})
K(z)^{-2}
I_{T}(z,h) \label{scigauging}
\end{equation} 
where
\begin{equation}
K(z_i)^{-1}=\prod_{n\ge 0} \left[ (1-q^{n+1})^{N-1} \prod_{i\neq j} (1-q^{n+1}\frac{z_i}{z_j})\right]\label{Kdef}.
\end{equation}
and  $z=\diag(z_1,\ldots,z_n)\in \SU(N)$ and $h\in H$.
\end{fact}
Here, in \eqref{scigauging}, the part $\prod_{i=1}^{N-1} {dz_i}/(2\pi \sqrt{-1} z_i)
\prod_{i\neq j} (1-{z_i}/{z_j})$ is the standard Haar measure of the Cartan torus of the $\SU(N)$ group manifold, 
and $K(z)^{-2}$ are the contributions from the other components of the vector multiplets. 
Here the formulas are stated for simplicity for $G=\SU(N)$, but it can be easily generalized to arbitrary gauge groups.

\subsection{The index of the $T_N$ theory and its cousins}
Now let us determine the index of the $T_N$ theory.
Our strategy is always the same, and we start by considering the index of the theory $S_{\SU(N)}\langle C_{g}\rangle$.
Almost by definition, this is the partition function of the 6d theory on $S^1\times S^3\times C_g$, with an R-symmetry background  preserving an appropriate number of supersymmetry. 
Now we use the basic fact of the 6d theory, and reduce along $S^1$ first. 
We have the \Nequals2 supersymmetric Yang-Mills theory with gauge group $\SU(N)$ on $S^3\times C$.
The 5d coupling constant is $8\pi^2 /g_5^2=1/R_6$. 

Now we have a Lagrangian and can perform the localization computation to get the partition function.
This computation was done in \cite{Fukuda:2012jr}\footnote{Strictly speaking, the background used in \cite{Fukuda:2012jr} is not the one that preserves \Nequals2 in 4d, but the one preserves \Nequals1. Still, from the study of \cite{Beem:2012yn} it is guaranteed solely by the supersymmetry that the index in this particular case equals the \Nequals2 Schur-limit index. The author thanks T. Kawano for discussions.}
The resulting theory is essentially the 2d $\SU(N)$ gauge theory on $C_g$, but the Kaluza-Klein modes along $S^3$ gives the dressing. 
The final answer is the 2d $q$-deformed Yang-Mills, with the parameter $q=\exp(-g_5^2/(4\pi R_{S^3}))=e^{-2\pi R_{S^1}/R_{S^3}}$.  The 2d $q$-deformed Yang-Mills was introduced in \cite{Alekseev:1994pa,Buffenoir:1994fh,Alekseev:1994au,Aganagic:2004js}.

This result allows us to write down the index $I_{g,n}$ of the theory $S_{\SU(N)}\langle C_{g,n}\rangle$ for the genus $g$ surface with $n$ full punctures as follows: \begin{equation}
I_{g,n}(a_i)=\sum_\lambda \frac{\prod_i N(a_i)\chi_\lambda(a_i)}{N_0^{n+2g-2}\chi_\lambda(q^{\rho})^{n+2g-2}} \label{scitmp}
\end{equation}
where $a_i\in \SU(N)$ is the flavor symmetry element for the $i$-th puncture, 
$\lambda$ runs over the irreducible representations of $\SU(N)$, 
$\chi_\lambda(a)$ is the character of the element $a$ in the representation $\lambda$,
and $q^{\rho} := (q^{(N-1)/2},q^{(N-3)/2},\cdots,q^{(1-N)/2})$.
Here we already took the limit where the area of $C_g$ is zero.

Here $N_0$ and $N(a)$ are renormalization factors the authors of \cite{Fukuda:2012jr} did not determine.
$N_0$ comes from the term of the form $c\int_C \sqrt{g}R = c(2-2g+n)$ in the action,
and $N(a)$ can come from the boundary term at the puncture. 
Both can be induced via renormalization, and the author does not know how to fix it by a direct computation.
We can still determine $N_0$ and $N(a)$ using the compatibility under the gluing and the complete closure, as shown below.

First, take two copies of $T_N=S_{\SU(N)}\langle C_{0,3}\rangle^\IR$, pick two punctures, and gauge them by an $\SU(N)$ vector multiplet.
The index of the resulting theory with $\SU(N)_a\times \SU(N)_b\times\SU(N)_c\times \SU(N)_d$ symmetry can be computed via Fact~\ref{fact:scigauging} and \eqref{scitmp}:
\begin{multline}
I(a,b;c,d)=\frac{1}{N!} \oint \prod_{i=1}^{N-1} \frac{dz_i}{2\pi \sqrt{-1} z_i}
\prod_{i,j} (1-\frac{z_i}{z_j})
K(z)^{-2}\\
\times \left(
\frac{N(a)N(b)N(z)}{N_0{}}\sum_\lambda \frac{\chi_\lambda(a)\chi_\lambda(b)\chi_\lambda(z)}{\chi_\lambda(q^{\rho})} \right) \\
\times\left(\frac{N(z)N(c)N(d)}{N_0{}}\sum_\lambda \frac{\chi_\lambda(z^{-1})\chi_\lambda(c)\chi_\lambda(d)}{\chi_\lambda(q^{\rho})}\right).
\end{multline}

But the resulting theory is $S_{\SU(N)}\langle C_{0,4}\rangle^\IR$,
and the index should have  the form  \eqref{scitmp} with $g=0$, $n=4$. 
To have this, we need \begin{equation}
\frac{1}{N!} \oint \prod_{i=1}^{N-1} \frac{dz_i}{2\pi \sqrt{-1} z_i}
\prod_{i,j} (1-\frac{z_i}{z_j})
K(z)^{-2}
N(z)^2 \chi_\lambda(z)\chi_\mu(z^{-1}) =\delta_{\mu\nu}.
\end{equation} 
Compare this equation with the orthogonality of the characters of the irreducible representations: \begin{equation}
\frac{1}{N!} \oint \prod_{i=1}^{N-1} \frac{dz_i}{2\pi \sqrt{-1} z_i}
\prod_{i,j} (1-\frac{z_i}{z_j})
 \chi_\lambda(z)\chi_\mu(z^{-1}) =\delta_{\mu\nu}.
\end{equation}
From this we see that the renormalization factors $N(z)$ we wanted to determine is given by  $N(z)=K(z)$. 

This factor $K(z)$ in the superconformal index can naturally be identified as the contribution from the conserved current multiplet in $\SU(N)$, including $\mu^{i=+1,0,-}$.
Note, for example, that the definition \eqref{Kdef} of $K(z)^{-1}$ involves a product over a basis of the adjoint of $\SU(N)$. 

Now, it is straightforward to obtain the superconformal index of theories with partially-closed punctures.
Originally, a full puncture has the contribution $K(a)\chi_\lambda(a)$ in the numerator of \eqref{scitmp}.
Let us set $\vev{\mu^+}=\rho_Y(\sigma^+)$.
The new $\SU(2)_R$ symmetry in the infrared is the diagonal combination of the original $\SU(2)_R$ 
and $\rho_Y(\SU(2))$. 
Denoting by $b$ an element of the flavor symmetry $G_Y$, 
this means that we perform the replacement 
\begin{equation}
a\to b q^{\rho_Y(\sigma_3)/2}
\end{equation}
in $\chi_\lambda(a)$ and $K(a)^{-1}$. 
The latter still contains the contributions from the Nambu-Goldstone modes that need to be removed: \begin{equation}
K(a) \xrightarrow{a\to bq^{\rho(\sigma_3)/2}} K_Y(b) \times (\text{contrib. from the NG modes}) 
\end{equation} where \begin{equation}
K_Y(b)^{-1}=\prod_d \prod_{w:\text{weights of $R_d$}} \prod_{n\ge 0} (1- q^{n+(d+1)/2} b^w).\label{genKdef}
\end{equation}   Here, we refined the decomposition \eqref{adjdecomp} of the adjoint under $\rho_Y(\SU(2))$ to the decomposition \begin{equation}
\mathbf{adj}=\bigoplus_d \underline{d} \otimes R_d
\end{equation} under $\rho_Y(\SU(2))\times G_Y$, where $\underline{d}$ is the $d$-dimensional irreducible representation of $\SU(2)$ as always, and $R_d$ is a representation of $G_Y$. 

In particular, when we completely close a puncture, we change a factor of $K(a)\chi_\lambda(a)$ in the numerator by $K_{[N]} \chi_\lambda(q^\rho)$. This should be equivalent to having one less puncture. Therefore, we should have \begin{equation}
N_0 = K_{[N]} = \prod_{d=2}^N \prod_{n\ge 0} (1-q^{d+n}). 
\end{equation} Now we completely determined the superconformal index.  Summarizing, we have 
\begin{fact}\label{fact:scigen}
The superconformal index in the Schur limit, of the 4d  theory $S_{\SU(N)}\langle C_{g,n}\rangle\{Y_1,\ldots,Y_n\}$  obtained from the 6d theory on a genus $g$ surface with punctures of type $Y_1$, \ldots, $Y_n$ is given by \begin{equation}
\sum_\lambda \frac{\prod_i K_{Y_i}(a_i)\chi_\lambda(a_i q^{\rho_{Y_i}})}{K_0^{n+2g-2}\chi_\lambda(q^{\rho})^{n+2g-2}}\label{generalSCIschur}
\end{equation}
where $a_i \in G_{Y_i}$, $q^{\rho_Y}:=q^{\rho_Y(\sigma^3)/2}$, $K_Y(a)$ is defined in \eqref{genKdef},
and $K_0:=K_{[N]}$, $q^{\rho}:=q^{\rho_{[N]}}$.
\end{fact}
This general result was first found in \cite{Gadde:2011ik}.

The theory $T_{Y_1,Y_2,Y_3}$ with a suitable choice of $Y_{1,2,3}$ can be a free hypermultiplet. 
In these cases, Fact~\ref{fact:scihyp} together with Fact~\ref{fact:scigen} implies an identity between an infinite sum and an infinite product. 
As examples we have the following equalities: \begin{fact}
The theory $T_{[1^N],[1^N],[N-1,1]}$  is a free theory of the bifundamental hypermultiplet. We then have the equality \begin{multline}
\prod_{u,i}\prod_{n\ge 0} \frac{1}{1-q^{n+1/2}(a_i/b_u)\alpha}
\frac{1}{1-q^{n+1/2}(b_u/a_i)/\alpha}
=\\
\frac{K(a)K(b)K_{[N-1,1]}(\alpha)}{K_0}
\sum_\lambda\frac{\chi_\lambda(a)\chi_\lambda(b) \chi_\lambda(q^{(N-2)/2}\alpha,\cdots,q^{(2-N)/2}\alpha,\alpha^{1-N})}
{\chi_\lambda(q^{(N-1)/2},\cdots,q^{(1-N)/2})}
\end{multline}   where \begin{equation}
K_{[N-1,1]}(\alpha)^{-1}=
\left[\prod_{d=1}^{N-1}\prod_{n\ge 0} (1-q^{d+n})\right]
\prod_{\pm}\prod_{n\ge 0} (1-q^{n+N/2}) \alpha^{\pm N}
\end{equation}  and $K(z)$ and $K_0$ were defined above. When $N=2$ the formula further simplifies and we have \begin{equation}
\prod_{\pm\pm\pm}\prod_{n\ge 0} \frac{1}{1-q^{n+1/2}a^\pm b^\pm c^\pm}
=\frac{K(a)K(b)K(c)}{K_0} \sum_d\frac{\chi_d(a)\chi_d(b)\chi_d(c)} {\chi_d(q^{1/2})} 
\end{equation}where $\chi_d(a)=a^{d-1}+a^{d-3}+\cdots+a^{1-d}$ is the $\SU(2)$ character in the $d$ dimensional irreducible representation. 
\end{fact} 
The proof of the case $N=2$ can be found in Appendix E of \cite{Gadde:2011uv}.

Also, the $T_3$ theory has an enhanced $E_6$ symmetry that is not manifest from the construction. Therefore, we have the following fact \begin{fact}
The $T_3$ theory has the $E_6$ symmetry. Therefore, its superconformal index has an expansion of the form \begin{equation}
\frac{K(a)K(b)K(c)}{K_0} \sum_\lambda\frac{\chi_\lambda(a)\chi_\lambda(b)\chi_\lambda(c)} {\chi_\lambda(q,1,q^{-1})}  = \sum_n q^n \chi_{R_n}(a_{1,2},b_{1,2},c_{1,2}) 
\end{equation}  where $R_n$ is a representation of $E_6$.
\end{fact}
Let us check this to $O(q^2)$. We have \begin{equation}
K(a)=1+\chi_8(a)q + O(q^2),\quad K_0=1+O(q^2)
\end{equation} 
and in the sum, only $\lambda=1$, $3$ and $\bar 3$ contribute. Then we have \begin{equation}
I(a,b,c)=1 + q(\chi_8(a)+\chi_8(b)+\chi_8(c)+\chi_3(a)\chi_3(b)\chi_3(c)
+\chi_{\bar 3}(a)\chi_{\bar 3}(b)\chi_{\bar 3}(c))+O(q^2).
\end{equation} We now see the decomposition \begin{equation}
\text{adj of $E_6$} = 8_A \oplus 8_B \oplus 8_C \oplus 3_A\otimes 3_B\otimes 3_C
\oplus  \bar 3_A\otimes \bar 3_B\otimes \bar 3_C
\end{equation} under $E_6 \supset \SU(3)_A\times \SU(3)_B\times \SU(3)_C$.

\subsection{Comments on further refinements}
In this review we wrote down the explicit formula \eqref{generalSCIschur} of the  superconformal index only for the special case $q=t$. As we argued above, this choice makes the superconformal index automatically independent of $p$. This choice is called the Schur limit. 

When $p=0$ with $q$ and $t$ generic, we still have an explicit formula generalizing \eqref{generalSCIschur}, obtained in \cite{Gadde:2011uv}. One crucial change is to replace the characters $\chi_\lambda(a)$ by the Macdonald polynomials $P_\lambda^{q,t}(a)$, that now depends on $q$ and $t$.  

An interesting subcase of the Macdonald limit is to take $p=q=0$. It is then conventional to use the variable $\tau=t^{1/2}$. The Macdonald polynomials reduce to the Hall-Littlewood polynomials $H_\lambda^{\tau}(a)$, and therefore this limit is called the Hall-Littlewood limit. This limit is particularly useful to study the Higgs branch of the theory, since it is known that, when the genus is $0$, the Hall-Littlewood limit of the superconformal index agrees with the Hilbert series of the Higgs branch.
  
The superconformal index with general three parameters $p$, $q$, $t$ can also be written in a form similar to \eqref{generalSCIschur}, by replacing the characters $\chi_\lambda(a)$ by suitable functions $\psi^{p,q,t}_\lambda(a)$ \cite{Gaiotto:2012xa} but the functions $\psi^{p,q,t}_\lambda(a)$ are not well understood, see e.g.~\cite{Razamat:2013qfa} for $G=\SU(2)$. 

The superconformal index in the Schur limit is also important from another point of view.  In \cite{Beem:2013sza}, it was shown that a 2d chiral algebra can be extracted from any  4d \Nequals2 superconformal theory by restricting operators to lie on a 2d plane in the 4d space, and that  the partition function of the vacuum module of this 2d chiral algebra equals the index in the Schur limit. The Schur index of the $T_N$ theory was further studied from this point of view in \cite{Beem:2013sza,Lemos:2014lua}.

The indices of  the theories $S_G\langle C_{g,n}\rangle$ for $G=D_n,E_n$ can of course be studied similarly; for explicit formulas, see \cite{Mekareeya:2012tn,Lemos:2012ph}. 
Also, the $T_{Y_1,Y_2,Y_3}$ theory for suitable choices of $Y_{1,2,3}$ is  a higher rank version of $E_n$ theories of Minahan and Nemeschansky and their superconformal indices are studied in detail, see e.g.~\cite{Gaiotto:2012uq,Keller:2012da,Hanany:2012dm}.

\section{Moduli spaces of supersymmetric vacua and chiral ring relations}\label{sec:moduli}
\subsection{Generalities on the moduli spaces of supersymmetric vacua}
A 4d \Nequals2 superconformal theory has a moduli space of supersymmetric vacua.  The part of the moduli space where $\SU(2)_R$ is unbroken is called the Coulomb branch, whereas the Higgs branch is where $\U(1)_R$ is unbroken.  The other parts are called the mixed branch. The chiral primary operators, in the \Nequals1 sense, that parametrize the Coulomb/Higgs branch is called the Coulomb/Higgs branch operators.  
Their scaling dimensions are fixed by the R-charges:\begin{fact}
The scaling dimension $\Delta$ of a Coulomb branch operator is $\Delta=r/2$,
and that of a Higgs branch operator is $\Delta=2I_3$.
\end{fact}

Higgs branch operators have intricate chiral ring relations, some of which will be discussed below.
As for the Coulomb  branch operators, a Lagrangian \Nequals2 gauge theory clearly does not have any relations among them, since it is a classic mathematical theorem that the gauge-invariant polynomials constructed from an adjoint operator is free of relations. 
For example, for $\SU(N)$, they are generated by $\tr \phi^k$ for $k=2,\ldots, N$.  
The Coulomb branches of the theories $S_G\langle C_{g,n}\rangle\{Y_1,\ldots,Y_n\}$ have been studied in detail, and no chiral ring relations have been found so far. Generalizing, we have
\begin{fact?}
The Coulomb branch operators are free of chiral ring relations. 
\end{fact?}

Assuming this, we have \cite{Shapere:2008zf}\begin{fact}\label{nv-intermsof-coulombdim}
The central charge $n_v$ and the spectrum of the Coulomb branch operators is related as follows: \begin{equation}
n_v = \sum_{u} (2\Delta(u)-1)
\end{equation}
where $u$ runs over the generators of the Coulomb branch operators. 
\end{fact}

Now, take a theory $\cQ\{G\}$ whose $G$ flavor symmetry can be conformally gauged. 
Then the Coulomb branch and the Higgs branch of the gauge theory $\cQ/G$ are related in a simple manner to those of the original theory $\cQ$:
\begin{fact}
The Coulomb branch operators of $\cQ/G$ consist of those of the theory $\cQ$ plus the gauge-invariant polynomials of the adjoint scalar $\phi$ in the \Nequals2 gauge multiplet.
When $G=\SU(N)$ those polynomials are $\tr\phi^k$, $k=2,\ldots,N$.
In particular, \begin{equation}
\dim_\bC \Coulomb(\cQ/G) = \dim_\bC\Coulomb(\cQ)+r
\end{equation}
where $r$ is the rank of $G$.
\end{fact}
\begin{fact}
The Higgs branch operators of $\cQ/G$ can be obtained by taking the Higgs branch operators of the theory $T$, setting $\mu^+=0$  where $\mu^+$ is the adjoint chiral operator in the $G$ current multiplet, and keeping only the $G$-invariant part.  As manifolds, this operation is often written as \begin{equation}
\Higgs(\cQ/G) = \Higgs(\cQ) \hkq G
\end{equation} and called the hyperk\"ahler quotient construction, introduced in \cite{Hitchin:1986ea}.
In particular, the dimension of the Higgs branch of $\cQ/G$ is given by 
\begin{equation}
\dim_\bH \Higgs(\cQ/G)=\dim_\bH \Higgs(\cQ) - (\dim G-\dim H)
\end{equation}
where $H$ is the unbroken gauge group at the generic point of the Higgs branch.
\end{fact}

Also, given a theory $\cQ\{G\}$ with flavor symmetry $G=\SU(N)$, we can partially close it by setting $\vev{\mu^+}=\rho_Y(\sigma^+)$ and removing the Nambu-Goldstone modes,
which are of the form $y=[\rho_Y(\sigma^+),x]$ where $x$ is an adjoint element of $\SU(N)$. 
Those $y$ that are not in this form can be singled out by imposing the constraint $[\rho_Y(\sigma^-),y]=0$.  Summarizing \cite{Moore:2011ee}:\begin{fact}\label{slicing}
The Higgs branch of the $\cQ\{Y\}$ theory obtained from the theory $\cQ$ by a partial closure is defined by \begin{equation}
\mu^+ \in S_Y, \qquad S_Y=\rho_Y(\sigma^+)+\{y\mid [\rho_Y(\sigma^-),y ]=0\}.
\end{equation}
The subspace $S_Y$ is called the Slodowy slice at $\rho_Y(\sigma^+)$.  In particular, 
\removevspaceafterequation
\begin{equation}
\dim_\bH \Higgs(\cQ\{Y\}) = \dim_\bH \Higgs(\cQ) - \frac12\dim_{\bC} O_Y/2.
\end{equation}
\end{fact} 

We do not yet have a good general way to understand the Coulomb branch of the partially closed theory $\cQ\{Y\}$, when $\cQ$ is \emph{not} a class S theory of type A. It would be desirable to have a method to understand this problem that applies to all 4d \Nequals2 theories.

\subsection{The moduli space of the $T_N$ theory}
Let us study the moduli space of the $T_N$ theory.  
Again the strategy is the same: we first consider the theory $S_{\SU(N)}\langle C_{g,0}\rangle$ for the  genus $g$ surface without punctures. Then we decompose it into copies of the $T_N$ theory and the contributions from the vector multiplets. 

To use the Lagrangian formalism, it is useful to compactify the 4d theory further on $S^1$. 
The resulting 3d theory  is  the 5d \Nequals2  theory with gauge group $\SU(N)$ on a genus $g$ surface $C$. 
Let us denote by $\phi_{1,2,3,4,5}$ the five adjoint fields in 5d. 
Due to the R-symmetry background, $\Phi_C=\phi_1+i\phi_2$ transforms as a one-form on $C$ with $\U(1)_R$ charge 2, while $\phi_{3,4,5}$ is an $\SU(2)_R$ triplet scalar on $C$. 
To make 3d \Nequals2 structure manifest, we combine $\phi_{3,4}$ into $\Phi_H=\phi_3+i\phi_4$. 
The other complex scalar in the hypermultiplet is a combination of $\phi_5$ and the scalar that is dual to the gauge field in 3d. 

The supersymmetric vacua correspond to the case when $\Phi_C$ and $\Phi_H$  commute, are holomorphic on $C$, considered up to complexified gauge transformations. 
The Coulomb branch corresponds to the situation $\Phi_C\neq 0$ while $\Phi_H=0$,
while the Higgs branch is the case where $\Phi_C=0$ and $\Phi_H\neq 0$. 

The Coulomb branch of the 3d theory is described by the so-called Hitchin system on $C$, and a nice review can be found in \cite{Neitzke:2014cja}. 
This is a hyperk\"ahler manifold whose complex dimension is twice that of the Coulomb branch in 4d.
The 4d Coulomb branch operators are encoded in terms of $\tr\Phi_C(z)^d$, which is a holomorphic $d$-differential on $C$, for $d=2,\ldots, N$. 
There are $(2d-1)(g-1)$ linearly-independent holomorphic $d$-differentials.
As $\tr\Phi_C(z)^d$ has $\U(1)_R$ charge $2d$, we find that there are $(2d-1)(g-1)$ Coulomb branch operators of scaling dimension $d$. 

As this theory is composed of $2(g-1)$ copies of the $T_N$ theory together with $3(g-1)$ copies of $\SU(N)$ \Nequals2 vector multiplet, one finds \begin{fact}\label{TnCoulomb}
The $T_N$ theory has $d-2$ Coulomb branch operators of scaling dimension $d$, for each $d=3,4,\ldots,N$.
\end{fact}
We can now combine this fact and Fact \ref{nv-intermsof-coulombdim} to derive $n_v$ of the $T_N$ theory. This nicely agrees with Fact \ref{TGnvnh}. 
Note also that we now deduced that the $T_N$ theory does not have any Coulomb branch operator of scaling dimension 2, $\U(1)_R$ charge 4. 
This means that the $T_N$ theory does not have any exactly marginal deformations.%
\footnote{The $T_N$ theory comes from a three-punctured sphere, that does not have any complex structure deformation. 
This fact alone does not guarantee that it does not have any exactly marginal deformations. 
Indeed, many examples are now known where a 4d theory obtained from a three-punctured sphere has exactly marginal deformations, see~\cite{Chacaltana:2012ch,Chacaltana:2013oka}.}

More precisely, the Seiberg-Witten curve of the 4d theory $S_{\SU(N)}\langle C_{g,0}\rangle$ is given by \begin{equation}
\det(\lambda-\Phi_C(z))=0\label{sw}
\end{equation} 
where $\lambda$ is the Seiberg-Witten one-form.  Note that $z$ is the coordinate of $C$ and $\lambda$ can be thought of as the coordinate along the cotangent direction of $T^*C$.
Then the equation \eqref{sw} determines an $N$-fold cover of the base $C$ embedded in $T^*C$.

As for the Higgs branch, we just give vevs to $\Phi_H$. They are zero-forms on $C$, so there are just $r=(N-1)$-dimensional Higgs branch, where we can put $\Phi_H$ to a  diagonal form. 
Note that this is independent of the genus $g$ of the curve.
To add $n$ full punctures, we consider the theory without punctures as the theory with $n$ completely closed punctures. 
When we completely close a full puncture, we lose the Higgs branch dimension by 
\begin{equation}
\frac12\dim_\bC O_{[N]}=\frac12(d-r)=\frac{N^2-N}{2}.
\end{equation}
Then, the dimension of the Higgs branch of the theory with $n$ punctures is $n(N^2-N)/2 + (N-1)$.  
Specializing to the case $g=0$, $n=3$, we find that the dimension of the Higgs branch of the $T_N$ theory is $3(d-r)/2 +r$. 
Note that this agrees with $n_h-n_v$ of the $T_N$ theory, as can be checked using Fact \ref{TGnvnh}.

This means that at the generic point on the Higgs branch of the $T_N$ theory, no free $\U(1)$ vector multiplet remains, because of the following analysis.
Recall the anomaly polynomial of the $T_N$ theory, which contains a term of the form $(n_v-n_h)c_1 p_1(TX)$. 
Giving a Higgs branch vev does not break $\U(1)_R$ symmetry, and therefore this term should be reproduced on a generic point on the Higgs branch. 
On a generic point, we just have free hypermultiplets whose number is given by the dimension of the Higgs branch, together with free $\U(1)$ vector multiplets. That $n_v-n_h$ agrees with the dimension of the Higgs branch then means that there is no free $\U(1)$ vector multiplet.

Now, take the theories $S_{\SU(N)}\langle C_{0,n}\rangle$ and $S_{\SU(N)}\langle C_{0,n'}\rangle$, and
connect them to form $S_{\SU(N)}\langle C_{0,n+n'-2}\rangle$.
By comparing the dimension of the Higgs branch before and after connecting them, 
we learn that the $G$ gauge symmetry is completely broken. 

Next, take a theory $S_{\SU(N)}\langle C_{g,n}\rangle$ with $n\ge 2$ punctures. 
Pick two punctures and connect them, to form the theory $S_{\SU(N)}\langle C_{g+1,n-2}\rangle$.
Again by comparing the dimension of the Higgs branch before and after connecting the punctures, we learn that the $G$ gauge symmetry is broken to a subgroup of rank $r=(N-1)$. 
This is in fact the Cartan subgroup $\U(1)^r$ of $G$.  
Repeating the procedure, we find that on the generic point on the Higgs branch of  the theory $S_{\SU(N)}\langle C_{g,n}\rangle$, we have $\U(1)^{rg}$ vector multiplets. 

This fact, when $n=0$, can be checked by another method. 
Without punctures, in 3d, we are giving a generic diagonalizable vev to $\Phi_H$. 
$\Phi_C$ should also be diagonal. 
Calling each diagonal entry $\Phi_C^{(i)}$ with $i=1,\ldots, N$ with $\sum_i \Phi_C^{(i)}=0$, we see that each $\Phi_C^{(i)}$ is a one-form and gives rise to $g$ $\U(1)$ vector multiplets in 4d. 
In total there are $(N-1)g$ $\U(1)$ multiplets in 4d, as we already found from slightly different perspective.
Summarizing, we have
\begin{fact}\label{fact:TnHiggs}
The Higgs branch of the $T_G$ theory has dimension $3(d-r)/2+r$. 
On a generic point on the Higgs branch, there remain no free vector multiplets. 
The action of the flavor symmetry $G^3$ is such that when a diagonal $G$ subgroup of $G^2$ is gauged by a $G$ gauge multiplet, the Cartan subgroup $\U(1)^r$ remains unbroken. 
\end{fact}

\subsection{Chiral ring relations of the $T_N$ theory}
Now let us discuss the chiral ring relations. The Coulomb branch operators do not have any nontrivial relation, so let us just discuss the Higgs branch chiral ring relations. They have been gradually being uncovered \cite{Benini:2009mz,Gadde:2013fma,Maruyoshi:2013hja,Yonekura:2013mya,Hayashi:2014hfa}, but we still do not have the complete understanding. The single most important one is
\begin{fact}
The operators $\mu^+_{A,B,C}$ in the adjoint of $\SU(N)_{A,B,C}$, in the $\SU(N)^3$ flavor symmetry multiplet, satisfy \begin{equation}
\tr (\mu^+_A)^k =\tr (\mu^+_B)^k =\tr (\mu^+_C)^k 
\end{equation}
for $k=2,\ldots,N$, and therefore for arbitrary $k$. 
From this reason we often drop the subscript $A,B,C$ in $\tr (\mu^+)^k$.
\end{fact} 

This was first understood via dualities involving Lagrangian gauge theories in \cite{Benini:2009mz}. 
Here we will use a version of the argument given in \cite{Yonekura:2013mya}, that is applicable to the $T_G$ theory for arbitrary $G$.  

\begin{figure}[h]
\[
\inc{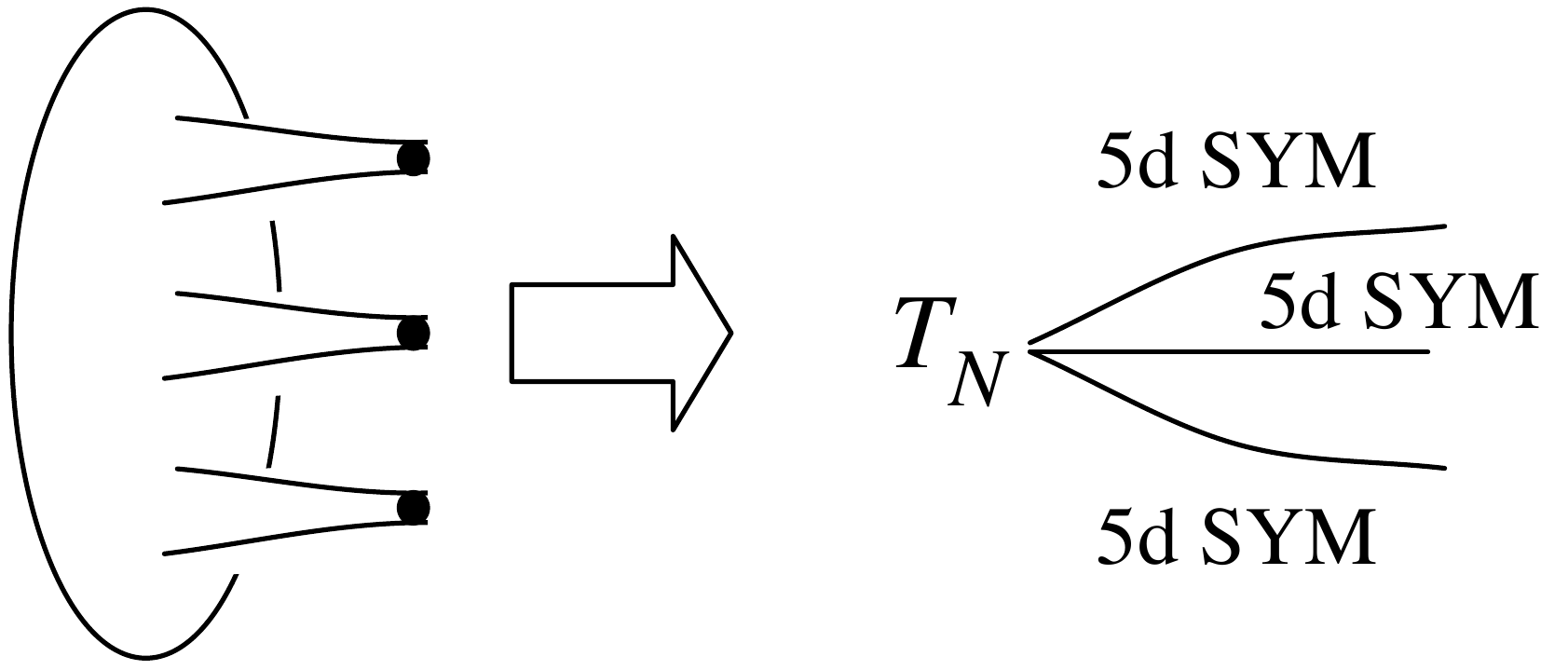}
\]
\caption{The $T_N$ theory with three segments of 5d super Yang-Mills attached.\label{fig:fubar}}
\end{figure}

We start from a three-punctured sphere with finite nonzero area, where the most of the area are concentrated at the tubes around the punctures, see Fig.~\ref{fig:fubar}. 
We can reduce this theory along the $S^1$ around three tubes. 
The result is the $T_G$ theory coupled to three segments of 5d \Nequals2 theory. 
When we take the strictly 4d limit, this setup just goes back to the $T_N$ theory.
One nice thing about this modification is that the operators $\mu^+_{A,B,C}$ are directly visible.
Indeed, three scalars $\phi_{3,4,5}$ of a segment  transform as a triplet under $\SU(2)_R$,
and three $G$ flavor symmetries act as gauge transformations at the boundaries of three segments. 
The $\mu^+_{A,B,C}$ fields for the $G_{A,B,C}$ flavor symmetry can then be identified 
as the boundary values of $\Phi_H=\phi_3 + i\phi_4$ at the ends of the segments. 

Now, let us compactify the entire setup further on another $S^1$.  
Then we can compare with the compactification of the 5d \Nequals2 super Yang-Mills on a Riemann surface we used in the last subsection. 
Comparing the two descriptions involves 3d mirror symmetry as detailed in \cite{Benini:2010uu}, 
but on a generic point on the Higgs branch where the dynamics is Abelian we can just  identify $\Phi_H$ used there and here. 
In particular, in a supersymmetric vacuum configuration, $\Phi_H$ should be holomorphic,
but $\Phi_H$ is a section of a trivial bundle, and
 therefore it is  a constant, up to  complexified gauge transformations. 
Therefore, three $\mu^+$ fields should be conjugate to each other, on a generic point on the Higgs branch, 
and we find $\tr (\mu^+_A)^k=\tr (\mu^+_B)^k=\tr(\mu^+_C)^k$ for arbitrary $k$.

Note that this argument breaks down when $\mu^+$ is non-generic due to the subtlety in the 3d mirror operation.  
For example, when we perform the partial closure, we often take $\vev{\mu^+_A}=\rho_Y(\sigma^+)\neq 0$ while trying to keep the other two vevs unchanged, $\vev{\mu^+_B}=\vev{\mu^+_C}=0$.

Let us move on to the analysis of Higgs branch operators other than $\mu^+$ in the flavor symmetry multiplet. 
The analysis is only applicable to the $T_N$ theory and not to the $T_G$ theory of general type. 
From the superconformal index in the Schur limit, it is easy to isolate the operators with lowest powers of $q$ in each $N$-ality of the $\SU(N)$ flavor symmetry, i.e.~the charge under $\bZ_N\subset \SU(N)$.
As the $K(a)$ factors only contain representations with zero $N$-ality, they come from the numerator $\chi_\lambda(a)\chi_\lambda(b)\chi_\lambda(c)$ with smallest possible $\lambda$ for each $N$-ality, 
and its power in $q$ is determined by the denominator $\chi_\lambda(q^{\rho})$.
In the sector with $N$-ality $k$, the  smallest possible $\lambda$ corresponds to the $k$-th antisymmetric tensor representation of $\SU(N)$ we denote by $\lambda=\wedge^k$.
As $\chi_{\wedge^k}(q^\rho) = q^{-k(N-k)/2} (1+O(q))$, 
we see that the leading contribution to the $N$-ality $k$ to the superconformal index is \begin{equation}
\tr_{H(S^3),\text{$N$-ality $k$}} (-1)^F q^{\Delta-I_3} abc  = q^{k(N-k)/2} \chi_{\wedge^k}(a)\chi_{\wedge^k}(b)\chi_{\wedge^k}(c) + \text{higher}.
\end{equation}
By studying the superconformal index, we can check that this contribution indeed comes from a scalar  operator with $\Delta=2I_3=k(N-k)$, transforming in $\wedge^k_A\otimes \wedge^k_B\otimes \wedge^k_C$. Summarizing, 
\begin{fact}
The $T_N$ theory has Higgs branch operators
\begin{itemize}
\item $Q_{aiu}$ with dimension $1(N-1)$,
\item   $Q_{[ab][ij][uv]}$ with dimension $2(N-2)$,  \ldots,
\item $Q_{[a_1\cdots a_k][i\cdots i_k][u_1\cdots u_k]}$ with dimension $k(N-k)$, \ldots,
\item $Q_{[a_{1}\cdots a_{N-1}][i_{1}\cdots i_{N-1}][u_{1}\cdots u_{N-1}]}$ with dimension $(N-1)1$,
\end{itemize} 
where $a$, $i$, $u$ are the indices for $\SU(N)_A$, $\SU(N)_B$, $SU(N)_C$, respectively. 
When $k>N/2$, it is often more convenient to raise the indices using epsilon symbols. 
For example, the last operator would become $\tilde Q^{aiu}$. 

In particular, for $N=2$, we just have dimension 1 operators $Q_{aiu}$,
and for $N=3$, we just have dimension 2 operators $Q_{aiu}$ and $\tilde Q^{aiu}$. 
\end{fact}
When $N=3$, the $T_3$ theory is the $E_6$ Minahan-Nemeschansky theory, for which the structure of the Higgs branch is known from different means. The fact above can then be checked \cite{Gaiotto:2008nz}.

The $Q$ operators introduced above, together with the operators $\mu^+_{A,B,C}$, are believed to generate the Higgs branch chiral ring. 
To study the chiral ring relations, we first note that
on a generic point on the Higgs branch, we can use flavor symmetry rotations to set \begin{equation}
\mu_A^+ = \mu_B^+ = \mu_C^+ = \diag(\mu_{1},\mu_{2},\ldots,\mu_N),
\qquad \sum \mu_i=0.\label{mudiag}
\end{equation} 
When we gauge a diagonal subgroup $\SU(N)$ of any two of $\SU(N)_A\times\SU(N)_B\times \SU(N)_C$, the $\U(1)^{N-1}$ Cartan subgroup should remain unbroken, as we saw in Fact~\ref{fact:TnHiggs}.
This means that the operator $Q_{[a_1\cdots a_k][i_1\cdots i_k][u_1\cdots u_k]}$ can be nonzero only when the multi-indices $a_1\cdots a_k$, $i_1\cdots i_k$ and $u_1\cdots u_k$ are the same up to the antisymmetry: 
\begin{equation}
Q_{[a_1\cdots a_k][i_1\cdots i_k][u_1\cdots u_k]} = q_{[a_1\cdots a_k]} \delta_{[a_1\cdots a_k],[i_1\cdots i_k],[u_1\cdots u_k]}\label{Qdiag}
\end{equation}
where we do not sum over the indices. 

This is consistent with a chiral ring relation we know from the superconformal index, that we describe now. 
Looking at the term of order $q^{(N+1)/2}$ in the superconformal index, we see that there is just one Higgs branch operator transforming in the trifundamental $\wedge_A \otimes \wedge_B \otimes \wedge_C$ with dimension $N+1$.\footnote{The coefficient of $q^{(N+1)/2}$ of the Schur limit index is $2$. One is a contribution from the descendant of $Q_{aiu}$ itself, and there is another that is a Higgs branch operator. The structure is clearer if we use the Hall-Littlewood limit instead, for which the descendants do not contribute.}
This means that there are two linear relations among $Q_{biu}(\mu^+_A)^b_a$, 
$Q_{aju}(\mu^+_B)^i_j$, and $Q_{aiv}(\mu^+_C)^v_u$. From the symmetry permuting three $\SU(N)$s, we find  that the relations are \begin{equation}
Q_{biu}(\mu^+_A)^b_a=Q_{aju}(\mu^+_B)^i_j=Q_{aiv}(\mu^+_C)^v_u.\label{Qmu}
\end{equation}
On a generic point on the Higgs branch where we have \eqref{mudiag}, these relations \eqref{Qmu} mean that $\mu_a Q_{abc}=\mu_b Q_{abc}=\mu_c Q_{abc}$. Therefore $Q_{abc}$ can be nonzero only when $a=b=c$.

Next, consider the two operators $Q_{[a[i(u} Q_{b]j]v)}$ and $Q_{[ab][ij][w(u]} (\mu^+_C)^w_{v)}$.
They both have scaling dimension $2N-2$ and transform in $\wedge_A^2 \otimes \wedge_B^2 \otimes \Sym^2_C$.
From the superconformal index, we can check that there is only one such Higgs branch operator, and therefore 
\begin{equation}
Q_{[a[i(u} Q_{b]j]v)} = Q_{[ab][ij][w(u]} (\mu^+_C)^w_{v)}.\label{QQ}
\end{equation}
On a generic point on the Higgs branch, we then find \begin{equation}
q_aq_b = q_{[ab]} (\mu_a-\mu_b), 
\end{equation}
where we use this relation to fix the relative normalizations of $Q_{aiu}$ and $Q_{[ab][ij][uv]}$.
From the consideration of the scaling dimensions and the remaining $S_N$ Weyl group action, it is natural to guess the general relation \begin{equation}
q_{a_1} \cdots q_{a_k} = q_{[a_1\cdots a_k]} \prod_{i<j} (\mu_{a_i}-\mu_{a_j}),\label{qs}
\end{equation} fixing the normalization of $Q_{[a_1\cdots a_k][i_1\cdots i_k][u_1\cdots u_k]}$.
In particular we have \begin{equation}
q_1\cdots q_N=\prod_{1\le i<j\le N}  (\mu_i -\mu_j).\label{qdet}
\end{equation}
This fixes the normalization of $Q_{aiu}$ itself.

Then we have $(N-1)$ complex degrees of freedom  in $\mu_1,\ldots \mu_N$ due to $\sum \mu_i=0$ 
and $(N-1)$ complex degrees of freedom in $q_1,\ldots q_N$ because the product $q_1\cdots q_N$ is fixed.
Thus we find $N-1$ hypermultiplet degrees of freedom in $\mu_i$ and $q_i$.  
In addition, to fix $\mu^+_{A,B,C}$ to the diagonal form, we have used $3(N^2-N)/2$  hypermultiplet degrees of freedom.
In total, there are $3(N^2-N)/2 +(N-1)$ hypermultiplet degrees of freedom, matching with the dimension of the Higgs branch of the $T_N$ theory. 
The \emph{non}-generic points on the Higgs branch is characterized by the vanishing of the discriminant of $\mu^+_{A,B,C}$, and its square root is given by  the right hand side of \eqref{qdet}.
From these analyisis, it is very likely that the $\mu$ operators for the three $\SU(N)$ flavor symmetries and the $Q$ operators generate all the Higgs branch operators.
Summarizing, 
\begin{fact?}
The operators $Q_{[a_1\cdots a_k][i_1\cdots i_k][u_1\cdots u_k]} $ together with the operators $\mu_{A,B,C}^+$ generate the Higgs branch chiral ring. Some of the relations involving the $Q$ operators are given in \eqref{Qmu} and \eqref{QQ}.
More generally, a necessary and sufficient condition for a candidate chiral ring relation is that it should be satisfied on generic points on the moduli space, i.e.~when we substitute  \eqref{mudiag}, \eqref{Qdiag}, \eqref{qs} and  \eqref{qdet} to the relation. 
\end{fact?}

The explicit forms of the  chiral ring relations for the $T_N$ theory, known in March 2015, can be found in Sec.~2 of \cite{Maruyoshi:2013hja} and in Appendix of \cite{Hayashi:2014hfa}. A different method to obtain the chiral ring relation by studying the 2d chiral algebra associated to the theory was pursued in \cite{Lemos:2014lua}.

\subsection{The moduli space of the partially closed theories}
As a final topic let us discuss the moduli space of the partially closed theories. 
As for the Higgs branch, we just apply the general method Fact~\ref{slicing} to the Higgs branch of the $T_N$ theory. 
We do not have much to say about the detailed structure of the chiral ring relations, but at least the dimension of the Higgs branch is easy to determine: we already computed the dimension for the theory $S_{\SU(N)}\langle C_{g,n}\rangle$ where all punctures are full. 
Then we perform the partial closures. We find:
\begin{fact}
The 4d theory $S_{\SU(N)}\langle C_{g,n}\rangle \{Y_1,\ldots,Y_n\}$, obtained by putting the 6d theory on a genus $g$ surface with $n$ punctures, labeled by $Y_1,\ldots,Y_n$, has a Higgs branch of dimension 
\removevspaceafterequation
\begin{equation}
\dim_\bH \Higgs(S_{\SU(N)}\langle C_{g,n}\rangle \{Y_1,\ldots,Y_n\}) =(r-1) + \sum_i  \frac12(d-r-\dim_\bC O_{Y_i}).
\end{equation}
\end{fact}
For $N=2$, $n=0$ and $g$ arbitrary, we see that the dimension is just 1. 
The precise form of the Higgs branch was determined in \cite{Hanany:2010qu} to be the asymptotically locally Euclidean space of type $D_{g+1}$.

In order to study the Coulomb branch, it is useful to revisit Fact~\ref{TnCoulomb} from a slightly different point of view. 
There, we obtained the dimension of the Coulomb branch of the $T_N$ theory by first counting the dimension of the Coulomb branch of the theory for genus $g$ surface without any puncture, and then decomposing them into the contributions from the $T_N$ theory and from the tubes. 
Combining them again, we find that the number of the Coulomb branch operators of scaling dimension $d$ of a theory for genus $g$ surface with $n$ full punctures is \begin{equation}
(2d-1)(g-1) + n (d-1).
\end{equation}
The contribution proportional to $(g-1)$ counts the dimension of the space of holomorphic $d$-differential on a genus $g$ surface, describing the degrees of freedom in $\tr \Phi_C(z)^d$.
The contribution proportional to $n$ can be accounted for by allowing $\tr \Phi_C(z)^d$ to have a pole at each full puncture, of order $d-1$.
This can be achieved if $\Phi_C(z)$ itself has a pole of the form \begin{equation}
\Phi_C(z) = \frac{edz}{z-z_0} + \text{regular} 
\end{equation}  where $z_0$ is the coordinate of a puncture, and $e$ is a generic nilpotent element.
Indeed, considering $\tr\Phi_C(z)^d$, we see that the term proportional to $(z-z_0)^{-d}$ drops out because $e$ is nilpotent, while the lower order terms are generically nonzero. 

Note that the full puncture has the type $[1^N]$, whereas this nilpotent element $e$ has the type $[N]$, which is the transpose of $[1^N]$.
In the other extreme, if a puncture at $z=z_0$ has the type $[N]$, it is equivalent to having no puncture at all, therefore the local form of the field $\Phi_C(z)$ is \begin{equation}
\Phi_C(z) = \frac{0dz}{z-z_0} + \text{regular},\label{0}
\end{equation}  to make no changes to the system. 
Now the residue $0$ is a nilpotent element of type $[1^N]$, and is given by the transpose of $[N]$.

In general, when the puncture at $z=z_0$ has the type $Y$, it is known that the field $\Phi_C(z)$ has the form \begin{equation}
\Phi_C(z) = \frac{e_{Y^t}dz}{z-z_0} + \text{regular},
\end{equation} where the residue $e_{Y^t}$ is a nilpotent element $e\in O_{Y^t}$ of type $Y^t$. 
 This can be argued in many ways, but one goes as follows. 
Consider partially closing a full puncture to the type  $Y=[n_1,n_2,n_3,\ldots,n_k]$, by setting $\vev{\mu_+}=J_Y$. 
This is compatible with the mass deformation associated to the original $\SU(N)$ symmetry of the full puncture given by \begin{equation}
m_Y=\diag(\underbrace{m_1,\ldots,m_1}_{n_1},\underbrace{m_2,\ldots,m_2}_{n_2},\ldots,\underbrace{m_k,\ldots,m_k}_{n_k}).
\end{equation} 
Under this deformation, $\Phi_C(z)$ should have a residue  at the puncture  of the form \begin{equation}
\Phi_C(z) = \frac{M_Y dz}{z-z_0} + \text{regular},
\end{equation} where $M_Y$ is a matrix whose eigenvalues agree with $m_Y$. 
This is due to the following:  the Seiberg-Witten curve is still given by \eqref{sw}, and the mass terms are given by the residues of $\lambda$. 
At $z=z_0$, the residues of $\lambda$ are clearly given by the eigenvalues of $M_Y$, and this should be identified with $m_Y$. 

Up to the action of $\SU(N)_C$,  we can always write $M_Y\sim m_Y+e$.
Here $A\sim B$ means two matrices are conjugate,
and $e$ is a nilpotent matrix within the Lie algebra of $G_Y$, the subgroup  of $\SU(N)$ left unbroken by the mass deformation $m_Y$.
This is just the standard Jordan decomposition of a matrix into the sum of a  diagonal matrix plus an upper-triangular matrix.

Within $G_Y$, the vev $\vev{\mu^+}=J_Y$ is the maximal possible one, i.e. we are completely closing the puncture within $G_Y$. 
Therefore, the residue of $\Phi_C(z)$ at $z=z_0$, when restricted to $G_Y$, should be zero, following the discussion around \eqref{0}. 
So we have $e=0$ and $M_Y \sim m_Y$. 
Now we turn off the eigenvalues $m_i$ of $m_Y$ to zero. 
Under this process, $M_Y$ does not necessarily tend to zero, but rather tend to a nilpotent matrix conjugate to $J_{Y^t}$, as explained in detail e.g.~in \cite{Gukov:2008sn}.
As an example, take $Y=[1^2]$. Then $M_Y \sim \diag(m,-m)$. 
 So take a one-parameter family of such $M_Y$ given by  \begin{equation}
M_Y=\begin{pmatrix}
m & 1 \\
0 & -m
\end{pmatrix}
\end{equation} and take $m\to 0$. We find \begin{equation}
M_Y \to \begin{pmatrix}
0 & 1 \\
0 & 0 
\end{pmatrix} = J_{[2]}.
\end{equation}
Summarizing, \begin{fact}
The Seiberg-Witten curve of the 4d theory $S_{\SU(N)}\langle C_{g,n}\rangle\{Y_1,\ldots,Y_n\}$, for the genus $g$ surface with $n$ punctures of type $Y_1,\ldots,Y_n$, is given by \begin{equation}
\det(\lambda-\Phi_C(z))=0\label{swgen}
\end{equation} where $\Phi_C(z)$ is a meromorphic differential on $C_{g,n}$, in the adjoint of $\SU(N)_\bC$, such that it has a pole at the $i$-th puncture at $z=z_i$ of the form \begin{equation}
\Phi_C(z)=\frac{e_{i}dz}{z-z_i}+\text{regular}
\end{equation} 
where $e_{Y_i^t}$ is a nilpotent element of type $Y_i^t$, i.e.~$e_i\in O_{Y_i^t}$.
\end{fact}

Given this, it is easy to count the dimension of the Coulomb branch. We can check that $\tr \Phi_C(z)^d$ at $z=z_i$ has a pole of order $p^d(Y_i)$, where \begin{equation}
p^d(Y) = d- \nu_d(Y), 
\end{equation} with the sequence $\nu_d$ defined by \begin{equation}
 (\nu_1,\nu_2,\ldots)=(\underbrace{1,\ldots,1}_{s_1},\underbrace{2,\ldots,2}_{s_2},\cdots)
\end{equation} where $Y^t = [s_1,s_2,\ldots]$. 
Then \begin{fact}
The number of Coulomb branch operators of dimension $d$ of the theory $S_{\SU(N)}\langle C_{g,n}\rangle\{Y_1,\ldots,Y_n\}$ for genus $g$ and $n$ punctures of type $Y_1$, \ldots, $Y_n$ is 
\removevspaceafterequation
\begin{equation}
(2d-1)(g-1) + \sum_i p^d(Y_i).
\end{equation}
\end{fact} 
For example, take the theory $T_{[1^N],[1^N],[N-1,1]}$. We have \begin{equation}
p([1^N])=(1,2,3,\ldots,N-1),\qquad
p([N-1,1])=(1,1,1,\ldots,1)
\end{equation} and therefore the number of Coulomb branch operators of this theory of scaling dimension $d$ is zero for all $d=2,\ldots,N$. 
This is as it should be, since this theory is a theory of free bifundamental hypermultiplets. 
Still, this analysis emphasizes an issue that was not clearly understood until several years ago, 
that a free hypermultiplets can still have a meaningful Seiberg-Witten curve \eqref{swgen}, 
that can be used in the analysis of the BPS geodesics, etc. 

Lastly let us note that these numbers $p^d$ satisfy two relations: \begin{equation}
\sum_d p^d(Y) = \dim_\bC O_{Y^t} /2, \qquad
\sum_d (2d-1)p^d(Y)=n_v(Y)
\end{equation}   where $n_v(Y)$ was introduced in \eqref{nvY}. 
The first means that the contribution from a puncture to the total dimension of the Coulomb branch is half the local degrees of freedom from the residue. 
This is reasonable, since the residue at a puncture contributes to the Coulomb branch dimension of the 3d theory, and the 4d Coulomb branch has half the dimension of that. 
The second equation means that the local contribution to $n_v$ and the local contribution to the number of the  Coulomb branch operators satisfy the sum rule Fact~\ref{nv-intermsof-coulombdim}. 

\section{Conclusions}\label{sec:conclusions}
In this article, we recalled the construction of the $T_N$ theory and its partially-closed cousins,
studied their flavor and conformal central charges,
determined their superconformal indices in the Schur limit,
and described the Coulomb and the Higgs branches of these theories. 

There are many topics on the $T_N$ theory the author could not cover in this review.  Let us at least list those properties where some works have been done:
\begin{itemize}
\item There are works on the line operators, the surface operators, and the boundary conditions of the $T_N$ theory, see e.g.~\cite{Ito:2011ea,Gang:2012yr,Xie:2013lca,Xie:2013vfa,Bullimore:2013xsa}.
See also the review \cite{Okuda:2014fja}.
\item There is also a 5d version \cite{Benini:2009gi,Bao:2013pwa} and a 3d version of the $T_N$ theory \cite{Benini:2010uu} and a closely related 3d theory called $T[\SU(N)]$ theory that captures the physics at a single puncture \cite{Gaiotto:2008ak}.
\item The Nekrasov partition function of the 5d $T_N$ theory can be computed by the topological string theory technique, and the way to take the 4d limit is now being analyzed in earnest \cite{Mitev:2014isa,Isachenkov:2014eya}.
\item The $T_N$ theory itself is \Nequals2 supersymmetric, but we can couple it to \Nequals1 gauge and matter multiplets and add superpotential terms, and study the strongly-coupled dynamics there. The combined system can be studied from the 6d point of view, by considering the R-symmetry background on the Riemann surface that only preserves \Nequals1 supersymmetry in 4d \cite{Bah:2011je,Bah:2012dg,Beem:2012yn,Gadde:2013fma,Maruyoshi:2013hja,Xie:2013gma,Bah:2013aha,Bonelli:2013pva,Xie:2013rsa,Agarwal:2013uga,Agarwal:2014rua,Giacomelli:2014rna,McGrane:2014pma,Xie:2014yya}.  
\item The holographic dual of the $T_N$ theory was already found in the original paper \cite{Gaiotto:2009gz} whereas the holographic dual for the $T_{D_n}$ theory was found in \cite{Nishinaka:2012vi}. 
The probes of the holographic duals for \Nequals1 theories were studied e.g.~in \cite{Bah:2013wda,Bah:2015fwa}.
\end{itemize}

Finally, we should remember that the $T_N$ theory has various non-supersymmetric correlation functions, whereas so far we only mentioned quantities that are protected either by topology (such as the anomaly) or by supersymmetry (such as the superconformal index or the Nekrasov partition function). 
Hopefully, one day, using a generalization of the superconformal bootstrap of \cite{Beem:2014zpa}, 
we might be able to compute the complete set of correlation functions of the $T_N$ theory.
But we are still far from that goal. 
The author hopes that we make steady progress in the next couple of years. 

\section*{Acknowledgements}
It is a pleasure for the author to thank D. Gaiotto, G. Moore and A. Neitzke; F. Benini, B. Wecht and D. Xie; O. Chacaltana and J. Distler; N. Mekareeya, J. Song; H. Hayashi, A. Gadde, K. Maruyoshi, W. Yan, and K. Yonekura for fruitful collaborations on the topic covered in this review.  The author would like to thank in particular K. Maruyoshi and K. Yonekura for carefully reading a draft version of this manuscript.
The work of the author is  supported in part by JSPS Grant-in-Aid for Scientific Research No. 25870159,
and in part by WPI Initiative, MEXT, Japan at IPMU, the University of Tokyo.

\bibliographystyle{ytphys}
\small\baselineskip=.9\baselineskip
\let\bbb\bibitem\def\bibitem{\itemsep1pt\bbb}
\bibliography{ref}

\end{document}